\documentclass[letterpaper,twocolumn,10pt]{article}
\usepackage{usenix2019_v3}
\usepackage[available,functional,reproduced]{usenixbadges}

\usepackage{geometry}
\newcommand{\sysname}{Clockwork\xspace}
\newcommand{\fakepara}[1]{\vspace{2mm}\noindent\textbf{#1}\hspace{5mm}}
\newcommand{\eg}{\textit{e.g.}\xspace}
\newcommand{\ie}{\textit{i.e.}\xspace}
\newcommand{\cf}{cf.\xspace}
\newcommand{\apriori}{\textit{a priori}\xspace}

\newcommand{\circone}{\protect\raisebox{-0.5pt}{\ding{192}}\xspace}
\newcommand{\circtwo}{\protect\raisebox{-0.5pt}{\ding{193}}\xspace}
\newcommand{\circthree}{\protect\raisebox{-0.5pt}{\ding{194}}\xspace}
\newcommand{\circfour}{\protect\raisebox{-0.5pt}{\ding{195}}\xspace}

\usepackage[utf8]{inputenc} %
\usepackage[table]{xcolor}
\usepackage{xspace} %
\usepackage{cite} %
\usepackage{paralist} %
\usepackage{graphicx} %
\usepackage{tikz} %
\usepackage{subcaption} %

\usepackage{tabu} %
\usepackage{longtable}
\usepackage{booktabs}
\usepackage{makecell} %
\usepackage{arydshln} %
\usepackage[usestackEOL]{stackengine} %
\usepackage{enumitem} %
\usepackage{pifont} %
\usepackage{stmaryrd} %
\usepackage{relsize} %
\usepackage[most]{tcolorbox} %
\tcbuselibrary{breakable}
\usepackage{adjustbox} %
\usepackage{array} %
\usepackage{multirow} %
\usepackage{verbatimbox} %
\usepackage{caption} %
\usepackage{layouts} %
\usepackage{bm} %
\usepackage[hang,flushmargin,bottom]{footmisc} %
\usepackage{scalerel}
\usepackage[super]{nth}
\usepackage{soul}
\usepackage{fancyvrb}

\usepackage{tabulary} %

\usepackage[subtle]{savetrees}

\usepackage[available,functional,reproduced]{usenixbadges}

\usepackage{epigraph} %
\usepackage{calc}
\newcommand{\mytextformat}{\rmshape\epigraphsize}

\let\originalepigraph\epigraph 
\renewcommand\epigraph[2]%
  {\setlength{\epigraphwidth}{\widthof{\mytextformat#2}}\originalepigraph{#1}{#2}}

\usepackage{stix} %

\usepackage{gnuplot-lua-tikz}

\usepackage{tikz} %
\usetikzlibrary{arrows.meta,bending,intersections,decorations.markings,shapes.geometric,decorations.pathreplacing,calc, trees, fadings}
\usetikzlibrary{fit,patterns,plotmarks,backgrounds, shapes.multipart}
\usetikzlibrary{positioning}

\usetikzlibrary{external}
\usepackage{titlesec} %
\titlespacing*{\section}{0pt}{2mm}{1mm}  %
\titlespacing*{\section}{0pt}{2mm}{1mm}  %
\titlespacing*{\subsection}{0pt}{2mm}{1mm}  %

\usepackage{setspace}  %
\usepackage{marginnote}
\usepackage{tabularx}

\usepackage{url}
\PassOptionsToPackage{colorlinks}{hyperref}
\PassOptionsToPackage{pdftex}{hyperref}

\usepackage[nameinlink,noabbrev]{cleveref} %
\DeclareCaptionLabelSeparator{forcespace}{~} %

\definecolor{figurecolor}{RGB}{22,90,220}
\definecolor{citecolor}{RGB}{198,81,19}
\usepackage{url}
\makeatletter
\g@addto@macro{\UrlBreaks}{\UrlOrds}
\makeatother

\def\Snospace~{\S{}}

\newcommand{\bestfont}[1]{\textsc{#1}}
\newcommand{\loadweights}{\bestfont{Load}\xspace}
\newcommand{\unloadweights}{\bestfont{Unload}\xspace}
\newcommand{\infer}{\bestfont{Infer}\xspace}
\newcommand{\copyinput}{\bestfont{Input}\xspace}
\newcommand{\exec}{\bestfont{Exec}\xspace}
\newcommand{\copyoutput}{\bestfont{Output}\xspace}
\newcommand{\earliest}{\texttt{earliest}\xspace}
\newcommand{\latest}{\texttt{latest}\xspace}

\hypersetup{ colorlinks=true, urlcolor=black, linkcolor=figurecolor, citecolor=citecolor, pdfstartview=FitH
        pdftitle={Serving DNNs like Clockwork: Performance Predictability from the Bottom Up},
        pdfauthor={Arpan Gujarati, Reza Karimi, Safya Alzayat, Wei Hao, Antoine Kaufmann, Ymir Vigfusson, Jonathan Mace}
        }

\newcommand\blfootnote[1]{%
  \begingroup
  \renewcommand\thefootnote{}\footnote{#1}%
  \addtocounter{footnote}{-1}%
  \endgroup
}

\usepackage{listings}

\definecolor{codegreen}{rgb}{0,0.6,0}
\definecolor{codegray}{rgb}{0.5,0.5,0.5}
\definecolor{codepurple}{rgb}{0.58,0,0.82}
\definecolor{backcolour}{rgb}{0.96,0.96,0.96}
\lstdefinestyle{mystyle}{
    backgroundcolor=\color{backcolour},   
    commentstyle=\color{codegreen},
    keywordstyle=\color{magenta},
    numberstyle=\tiny\color{codegray},
    stringstyle=\color{codepurple},
    basicstyle=\ttfamily\footnotesize,
    breakatwhitespace=false,         
    breaklines=true,                 
    captionpos=b,                    
    keepspaces=true,                 
    numbers=left,                    
    numbersep=5pt,                  
    showspaces=false,                
    showstringspaces=false,
    showtabs=false,                  
    tabsize=2
}
\begin{document}

\title{\Large \bf Serving DNNs like \sysname{}: Performance Predictability from the Bottom Up}

\author{
\begin{tabular}{c c}
{\rm Arpan Gujarati\footnotemark[1]} %
& \qquad
{\rm Reza Karimi\footnotemark[1]}\\[-0.5mm]
Max Planck Institute for Software Systems
& \qquad
Emory University\\[8pt]
{\rm Safya Alzayat} 
& \qquad
{\rm Wei Hao}\\[-0.5mm]
Max Planck Institute for Software Systems
& \qquad
Max Planck Institute for Software Systems\\[8pt]
{\rm Antoine Kaufmann} 
& \qquad
{\rm Ymir Vigfusson}\\[-0.5mm]
Max Planck Institute for Software Systems
& \qquad
Emory University \\[8pt]
\multicolumn{2}{c}{\rm Jonathan Mace} \\[-0.5mm]
\multicolumn{2}{c}{Max Planck Institute for Software Systems}
\end{tabular}
}

\maketitle

\begin{abstract}
Machine learning inference is becoming a core building block for interactive web applications.
As a result, the underlying model serving systems on which these applications depend must consistently meet low latency targets.
Existing model serving architectures 
use well-known reactive techniques to alleviate common-case sources of latency,
but cannot effectively curtail tail latency caused by unpredictable execution times. %
Yet the underlying execution times are not fundamentally unpredictable---on the contrary we observe that 
inference using Deep Neural Network (DNN) models has deterministic performance.

Here, starting with the predictable execution times of individual DNN inferences, we adopt a principled design methodology to successively build a fully distributed model serving system that achieves predictable end-to-end performance.
We evaluate our implementation, \sysname{}, using production trace workloads, and show that \sysname{} can support thousands of models while simultaneously meeting 100\,ms latency targets for 99.9999\% of requests.
We further demonstrate that \sysname{} exploits predictable execution times to achieve tight request-level service-level objectives (SLOs) as well as a high degree of request-level performance isolation.
\end{abstract}

\blfootnote{\hspace{3mm}\small * Equal contribution}

\section{Introduction}
\label{sec:intro}

With the proliferation of machine learning (ML), model inferences are now not only commonplace but increasingly on the critical path of web requests~\cite{wu2019machine,hazelwood2018applied}.
Inference requests are handled by underlying model serving services~\cite{olston2017tensorflowserving,crankshaw2017clipper,romero2019infaas,googleaiplatform} responsible for supporting scores of different pre-trained ML models (including personalized models and experimental A/B tests), ideally at low latency, high throughput, and low cost. 
These are demanding goals to meet at scale---Facebook alone serves over 200 \emph{trillion} inference requests each day \cite{mattson2020mlperf}.
Furthermore, at least 100 companies are creating hardware chips for accelerated ML inference \cite{mattson2020mlperf},
which underscores the high stakes in this industry.

Yet significant software bottlenecks continue to hamper the efficient utilization of hardware accelerators, such as GPUs, for high-performance model serving.
Consider an inference request passing through a model serving system. 
The request has an inherent deadline after which the answer ceases to be useful to the end-user,
and so the system should seek to bound the latency of the request, or even provide service level objectives (SLOs) for consistently achieving low tail latency.
The canonical approach for building such a low-latency system is to reduce potential wait times for resources through  over-provisioning, since a larger pool of available resources makes
it more likely to find a resource on which a pending request can be immediately scheduled.
Increased resource provisioning, however, comes at the expense of efficiency and utilization. 

Existing systems fundamentally assume that the constituent system components have \emph{unpredictable}  latency performance~\cite{romero2019infaas,crankshaw2017clipper}. 
Moreover, the best-effort techniques employed to tolerate such variability, such as fair queuing, further cascade the unpredictability to other system components and propagate tail latency to higher layers.
While some performance volatility of a model serving system is due to external factors, such as a bursty or skewed workload, much variability in execution times stems from design decisions internal to the service, ranging from caching decisions over conditional branching behavior to concurrency from other processes, the OS, and the hypervisor. 
The challenge, then, is to tame the internal unpredictability.

In this paper, we present the design and implementation of \sysname{}, a distributed system for serving models with predictable performance.
With an explicit focus on the ubiquitous deep neural network (DNNs) architectures
we first show that DNN inference is fundamentally a deterministic sequence of mathematical operations that has a predictable execution time on a GPU.
To leverage this observation in designing a responsive model serving system,
our approach is to preserve predictability wherever possible by \emph{\textbf{consolidating choice}}: eschewing reactive and best-effort mechanisms and centralizing all resource consumption and scheduling decisions.
\sysname{} will only execute an inference request if it is confident that the request can meet its latency SLO.
To support such proactive scheduling, \sysname{} is composed of \emph{workers} that each handle one or more GPUs, and a centralized \emph{controller} that schedules requests. 
Each \sysname{} worker, responsible for the exclusive model loading and inference execution on the GPUs, achieves predictable performance. 
If a worker cannot execute a particular schedule, because of external factors, the request is immediately aborted and the worker resumes execution of the next request at the specified time.
The \sysname{} controller manages the resources of each worker and maintains a minimal advance schedule for the worker's operations, including model placement and replication.

We have implemented \sysname{} in C\texttt{++} and evaluated it using a wide range of DNN models on production workload traces.  In comparison to Clipper~\cite{crankshaw2017clipper} and INFaaS~\cite{romero2019infaas}, two prior model serving systems, \sysname more effectively meets latency goals while providing comparable or better goodput.  \sysname more effectively shares resources between different models, and scales to thousands of models per worker.  For realistic workloads comprising unpredictable, bursty, and cold-start clients, \sysname consistently meets low-latency response times of under 100ms.

The main contributions of this paper are as follows:

\begin{compactitem}
\item We demonstrate that predictability is a fundamental trait of DNN inference that can be exploited to build a predictable model serving system.
\item We propose a system design approach, \emph{consolidating choice}, to preserve predictable responsiveness in a larger system comprised of components with predictable performance.
\item We present the design and implementation of \sysname, a distributed model serving system that mitigates tail latency of DNN inference from the bottom up.
\item We report from an experimental evaluation on \sysname to show that the system supports thousands of models concurrently per GPU and substantially mitigates tail latency, even while supporting tight latency SLOs.  \sysname achieves close to ideal goodput even under overload, with unpredictable and bursty workloads, and with many contending users.
\end{compactitem}

\section{Background and Motivation}
\label{sec:motivation}

\fakepara{The state of machine learning.}
The meteoric rise of applications driven by machine learning (ML), ranging from computer vision \cite{zhang2020resnest,guo2020gluoncv} to ad-targeting \cite{dalessandro2014scalable, agarwal2014laser} to virtual assistants \cite{campagna2017almond, sokol2018glass}, has prompted significant 
interest into making both ML training and inference faster. 
These efforts have targeted the underlying ML models, hardware accelerators, and software infrastructure.
Chief among the ML modeling approaches are \emph{deep neural networks} (DNNs), which are composed of multiple layers of artificial neurons tuned through non-linear convolution and pooling operations~\cite{goodfellow2016deep}. %

A plethora of specialized hardware are being developed and deployed for ML training and inference~\cite{mattson2020mlperf}, such as ASIC and FPGA chips, Google's TPUs~\cite{jouppi2017datacenter}, and Facebook's Big Basin~\cite{hazelwood2018applied} chips.
The dominant machine learning hardware in data centers, however, is the GPU, representing a third of the global market in 2020~\cite{chipmarket}, and will be our focus here.

\begin{figure}
\centering
\includegraphics[page=5,width=\columnwidth, trim=0 275 90 0,clip]{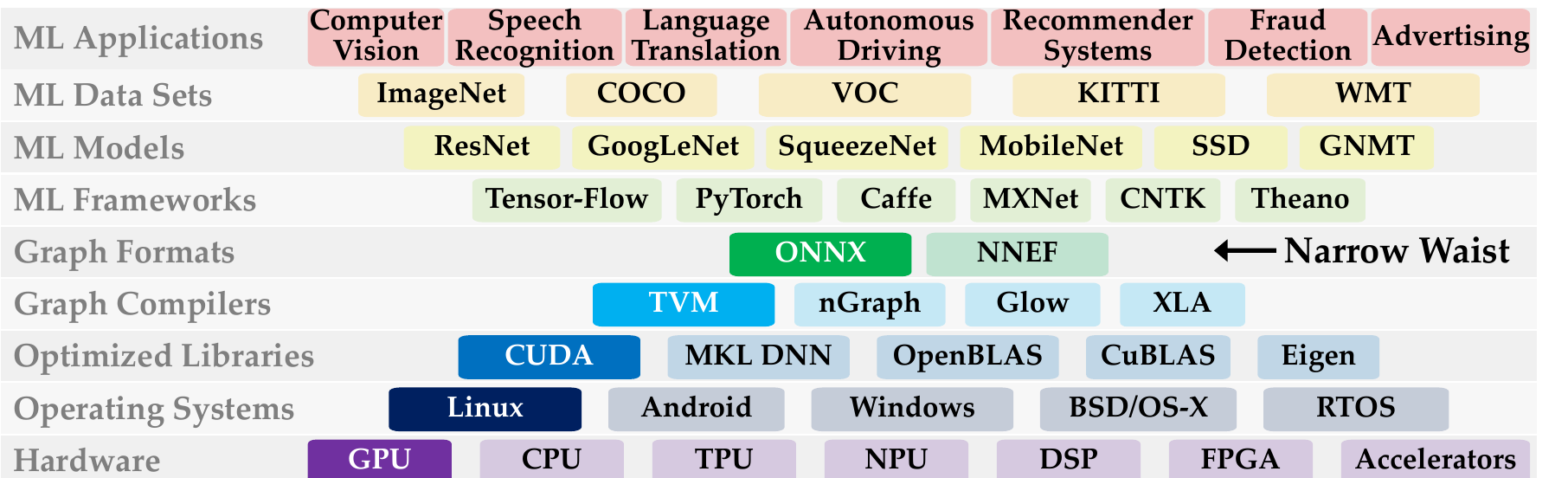}%
\caption{Model serving targets the narrow waist of the ML software stack (adapted from Reddi \emph{et al.}~\cite{mattson2020mlperf}). \sysname{} targets the shaded blocks on the left.
}%
\label{fig:narrowwaist}%
\end{figure}

Interposed between the emerging DNN applications and hardware accelerators, an ecosystem of ML software frameworks is flourishing. 
\autoref{fig:narrowwaist} displays several prominent projects in today's ML software stack.
Layered protocol stacks in complex systems and competitive environments tend to evolve into hourglass-shaped architectures~\cite{akhshabi2011evolution}. 
We are witnessing the ONNX and NNEF graph
exchange formats for DNNs~\cite{onnx,nnef} emerging as the ``narrow waist'' of the ML stack, 
acting as an interface between high-level ML model development and low-level software and hardware concerns.

\fakepara{Model serving.}
Operators increasingly deploy machine learning on the critical path of nascent interactive applications~\cite{wu2019machine}.
This has elevated machine learning inference to separate, managed \emph{model serving} services~\cite{romero2019infaas,crankshaw2017clipper,googleaiplatform}.
From the vantage point of an operator, the model serving users (customers or internal applications) upload their pre-trained DNN ahead of time (the natural format for which is ONNX/NNEF).  Their applications can then submit \emph{inference requests} to an API.
The model serving back-end manages the users' models and the hardware accelerator resources, and provides timely responses to inference requests.
Upon receiving an inference request, it loads the appropriate model into hardware if not already loaded, runs the DNN on the input, and returns the resulting output to the user. %
Model serving has similar concerns to other datacenter services~\cite{adya2016slicer}: it multiplexes workloads of different users concurrently and load balances requests across multiple workers and GPU hardware accelerators.

\fakepara{Low-latency inference.}
Model serving users require a timely response to their queries.
Most cloud and data center services have \emph{service-level objectives} (SLOs) that codify the performance that clients can expect from the service~\cite{srebook}.  The most common type is a \emph{latency SLO}, which specifies the service's acceptable request latencies, typically on the order of milliseconds~\cite{hermann2017michelangelo,jouppi2017datacenter,cheeyau2017zendesk}.  For example, a latency SLO might specify a 10ms average response time, or a 40ms 99$^\text{th}$ percentile response time, or both.  If a service fails to meet its SLOs -- for example, by being too slow for too many requests -- the service provider may risk a penalty.

Model serving further operates under hard cost constraints. 
Specialized ML hardware is necessary to achieve interactive latencies~\cite{jouppi2017datacenter}, but it is comparatively expensive to procure and operate, and must thus be used efficiently~\cite{schwartz2019green,strubell2019energy}.
Existing model serving systems achieve efficient inferences for specific heavily used models by dedicating them entire GPUs and using copious batching~\cite{jouppi2017datacenter}.
However, many use cases cannot justify dedicated hardware resources: applications with insufficient request volume; specialization (\eg location-specific search or language-to-language translation); and experimentation (\eg retrained models and A/B testing)~\cite{simon2018elasticinference}.
Efficiently serving models with low request rates requires a large number of models to share accelerators; no existing model serving system supports this.

While it is already difficult for model serving operators to meet latency SLOs under these constraints, the bigger challenge lies in minimizing \emph{tail latency}, the insidious bane of interactive performance.
Numerous sources of latency variability in complex individual~\cite{li2014tales} and distributed~\cite{dean2013tail,ousterhout2015making} systems have been identified and studied, including out-of-order scheduling, interference from concurrency, power saving modes, and network queuing delays. 

The crux of tail latency lies in performance variability of both the constituent
system/network components and the encompassing architecture.
To tame it, the system designer can either seek to (quoting Dean and Barrosso~\cite{dean2013tail}) ``\emph{create a predictably responsive whole out of less-predictable parts}'', or 
to expend significant effort to systematically unshroud and mitigate the performance variability of these underlying components. 
To meet tight tail-latency SLOs under resource constraints, the latter approach is necessary.

\fakepara{Observation: DNN inference is predictable.}
We observe that DNN executions exhibit negligible latency variability,
a result both intuitive in concept --- DNN inferences involve no conditional branches --- and demonstrable in practice.
Although we describe our observations in the context of GPU execution, they extend to other accelerators such as TPUs, and also to CPU execution where appropriate.

Conceptually, a DNN inference is a fully deterministic execution. 
Each DNN inference request  carries a fixed-size \emph{input} tensor argument; in practical terms this is a statically-sized array of bytes.  A worker receives this input over the network into main memory.  To execute on a GPU, the input is copied from main memory to GPU memory over the PCIe interconnect.  The DNN is then executed on the GPU.  Abstractly, a DNN is a pre-defined sequence of tensor multiplications and activation functions.  Concretely, the DNN code applies these operations to the input tensor one-at-a-time to transform the input into an output.  DNN code lacks conditional branching; input choices such as batching size and RNN sequence length are specified ahead of time as parameters. The output is also a statically-sized array of bytes, and it is copied from GPU memory back to main memory over the PCIe interconnect. 

\begin{figure}[t]%
\begin{subfigure}[t]{3.75cm}%
\tikzsetnextfilename{throughput_microbenchmark_tail}%
\input{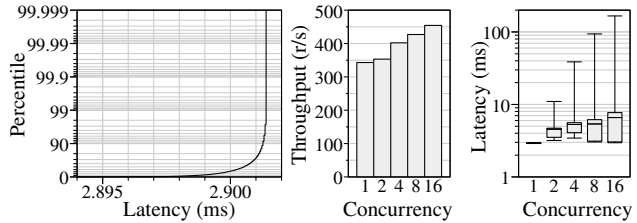}%
\caption{CDF of 1-thread latency}%
\label{fig:dnninferencetail}%
\end{subfigure}%
\begin{subfigure}[t]{4.67cm}%
\tikzsetnextfilename{throughput_microbenchmark_throughput}%
\begin{tikzpicture}[gnuplot]
\tikzset{every node/.append style={font={\fontsize{7.0pt}{8.4pt}\selectfont}}}
\path (0.000,0.000) rectangle (2.335,3.000);
\gpcolor{rgb color={0.800,0.800,0.800}}
\gpsetlinetype{gp lt border}
\gpsetdashtype{gp dt solid}
\gpsetlinewidth{1.00}
\draw[gp path] (0.887,0.690)--(2.240,0.690);
\gpcolor{color=gp lt color border}
\draw[gp path] (0.887,0.690)--(0.833,0.690);
\node[gp node right] at (0.936,0.690) {$0$};
\gpcolor{rgb color={0.800,0.800,0.800}}
\draw[gp path] (0.887,0.912)--(2.240,0.912);
\gpcolor{color=gp lt color border}
\draw[gp path] (0.887,0.912)--(0.860,0.912);
\gpcolor{rgb color={0.800,0.800,0.800}}
\draw[gp path] (0.887,1.134)--(2.240,1.134);
\gpcolor{color=gp lt color border}
\draw[gp path] (0.887,1.134)--(0.833,1.134);
\node[gp node right] at (0.936,1.134) {$100$};
\gpcolor{rgb color={0.800,0.800,0.800}}
\draw[gp path] (0.887,1.356)--(2.240,1.356);
\gpcolor{color=gp lt color border}
\draw[gp path] (0.887,1.356)--(0.860,1.356);
\gpcolor{rgb color={0.800,0.800,0.800}}
\draw[gp path] (0.887,1.578)--(2.240,1.578);
\gpcolor{color=gp lt color border}
\draw[gp path] (0.887,1.578)--(0.833,1.578);
\node[gp node right] at (0.936,1.578) {$200$};
\gpcolor{rgb color={0.800,0.800,0.800}}
\draw[gp path] (0.887,1.800)--(2.240,1.800);
\gpcolor{color=gp lt color border}
\draw[gp path] (0.887,1.800)--(0.860,1.800);
\gpcolor{rgb color={0.800,0.800,0.800}}
\draw[gp path] (0.887,2.021)--(2.240,2.021);
\gpcolor{color=gp lt color border}
\draw[gp path] (0.887,2.021)--(0.833,2.021);
\node[gp node right] at (0.936,2.021) {$300$};
\gpcolor{rgb color={0.800,0.800,0.800}}
\draw[gp path] (0.887,2.243)--(2.240,2.243);
\gpcolor{color=gp lt color border}
\draw[gp path] (0.887,2.243)--(0.860,2.243);
\gpcolor{rgb color={0.800,0.800,0.800}}
\draw[gp path] (0.887,2.465)--(2.240,2.465);
\gpcolor{color=gp lt color border}
\draw[gp path] (0.887,2.465)--(0.833,2.465);
\node[gp node right] at (0.936,2.465) {$400$};
\gpcolor{rgb color={0.800,0.800,0.800}}
\draw[gp path] (0.887,2.687)--(2.240,2.687);
\gpcolor{color=gp lt color border}
\draw[gp path] (0.887,2.687)--(0.860,2.687);
\gpcolor{rgb color={0.800,0.800,0.800}}
\draw[gp path] (0.887,2.909)--(2.240,2.909);
\gpcolor{color=gp lt color border}
\draw[gp path] (0.887,2.909)--(0.833,2.909);
\node[gp node right] at (0.936,2.909) {$500$};
\node[gp node center] at (1.113,0.582) {1};
\node[gp node center] at (1.338,0.582) {2};
\node[gp node center] at (1.564,0.582) {4};
\node[gp node center] at (1.789,0.582) {8};
\node[gp node center] at (2.015,0.582) {16};
\draw[gp path] (0.887,2.909)--(0.887,0.690)--(2.240,0.690)--(2.240,2.909)--cycle;
\node[gp node center] at (1.546,0.270) {\relsize{1}{Concurrency}};
\node[gp node center,rotate=-270] at (0.241,1.799) {\relsize{1}{Throughput (r/s)}};
\gpfill{rgb color={0.933,0.933,0.933}} (1.000,0.690)--(1.226,0.690)--(1.226,2.213)--(1.000,2.213)--cycle;
\draw[gp path] (1.000,0.690)--(1.000,2.212)--(1.225,2.212)--(1.225,0.690)--cycle;
\gpfill{rgb color={0.933,0.933,0.933}} (1.225,0.690)--(1.452,0.690)--(1.452,2.258)--(1.225,2.258)--cycle;
\draw[gp path] (1.225,0.690)--(1.225,2.257)--(1.451,2.257)--(1.451,0.690)--cycle;
\gpfill{rgb color={0.933,0.933,0.933}} (1.451,0.690)--(1.677,0.690)--(1.677,2.475)--(1.451,2.475)--cycle;
\draw[gp path] (1.451,0.690)--(1.451,2.474)--(1.676,2.474)--(1.676,0.690)--cycle;
\gpfill{rgb color={0.933,0.933,0.933}} (1.676,0.690)--(1.903,0.690)--(1.903,2.586)--(1.676,2.586)--cycle;
\draw[gp path] (1.676,0.690)--(1.676,2.585)--(1.902,2.585)--(1.902,0.690)--cycle;
\gpfill{rgb color={0.933,0.933,0.933}} (1.902,0.690)--(2.128,0.690)--(2.128,2.707)--(1.902,2.707)--cycle;
\draw[gp path] (1.902,0.690)--(1.902,2.706)--(2.127,2.706)--(2.127,0.690)--cycle;
\draw[gp path] (0.887,2.909)--(0.887,0.690)--(2.240,0.690)--(2.240,2.909)--cycle;
\gpdefrectangularnode{gp plot 1}{\pgfpoint{0.887cm}{0.690cm}}{\pgfpoint{2.240cm}{2.909cm}}
\end{tikzpicture}%
%
\tikzsetnextfilename{throughput_microbenchmark_latency}%
\begin{tikzpicture}[gnuplot]
\tikzset{every node/.append style={font={\fontsize{7.0pt}{8.4pt}\selectfont}}}
\path (0.000,0.000) rectangle (2.335,3.000);
\gpcolor{rgb color={0.800,0.800,0.800}}
\gpsetlinetype{gp lt border}
\gpsetdashtype{gp dt solid}
\gpsetlinewidth{1.00}
\draw[gp path] (0.758,0.690)--(2.240,0.690);
\gpcolor{color=gp lt color border}
\draw[gp path] (0.758,0.690)--(0.668,0.690);
\node[gp node right] at (0.771,0.690) {1};
\gpcolor{rgb color={0.800,0.800,0.800}}
\draw[gp path] (0.758,1.654)--(2.240,1.654);
\gpcolor{color=gp lt color border}
\draw[gp path] (0.758,1.654)--(0.668,1.654);
\node[gp node right] at (0.771,1.654) {10};
\gpcolor{rgb color={0.800,0.800,0.800}}
\draw[gp path] (0.758,2.619)--(2.240,2.619);
\gpcolor{color=gp lt color border}
\draw[gp path] (0.758,2.619)--(0.668,2.619);
\node[gp node right] at (0.771,2.619) {100};
\gpcolor{rgb color={0.800,0.800,0.800}}
\draw[gp path] (0.758,0.690)--(2.240,0.690);
\gpcolor{color=gp lt color border}
\draw[gp path] (0.758,0.690)--(0.668,0.690);
\gpcolor{rgb color={0.800,0.800,0.800}}
\draw[gp path] (0.758,0.980)--(2.240,0.980);
\gpcolor{color=gp lt color border}
\draw[gp path] (0.758,0.980)--(0.713,0.980);
\gpcolor{rgb color={0.800,0.800,0.800}}
\draw[gp path] (0.758,1.150)--(2.240,1.150);
\gpcolor{color=gp lt color border}
\draw[gp path] (0.758,1.150)--(0.713,1.150);
\gpcolor{rgb color={0.800,0.800,0.800}}
\draw[gp path] (0.758,1.271)--(2.240,1.271);
\gpcolor{color=gp lt color border}
\draw[gp path] (0.758,1.271)--(0.713,1.271);
\gpcolor{rgb color={0.800,0.800,0.800}}
\draw[gp path] (0.758,1.364)--(2.240,1.364);
\gpcolor{color=gp lt color border}
\draw[gp path] (0.758,1.364)--(0.713,1.364);
\gpcolor{rgb color={0.800,0.800,0.800}}
\draw[gp path] (0.758,1.440)--(2.240,1.440);
\gpcolor{color=gp lt color border}
\draw[gp path] (0.758,1.440)--(0.713,1.440);
\gpcolor{rgb color={0.800,0.800,0.800}}
\draw[gp path] (0.758,1.505)--(2.240,1.505);
\gpcolor{color=gp lt color border}
\draw[gp path] (0.758,1.505)--(0.713,1.505);
\gpcolor{rgb color={0.800,0.800,0.800}}
\draw[gp path] (0.758,1.561)--(2.240,1.561);
\gpcolor{color=gp lt color border}
\draw[gp path] (0.758,1.561)--(0.713,1.561);
\gpcolor{rgb color={0.800,0.800,0.800}}
\draw[gp path] (0.758,1.610)--(2.240,1.610);
\gpcolor{color=gp lt color border}
\draw[gp path] (0.758,1.610)--(0.713,1.610);
\gpcolor{rgb color={0.800,0.800,0.800}}
\draw[gp path] (0.758,1.654)--(2.240,1.654);
\gpcolor{color=gp lt color border}
\draw[gp path] (0.758,1.654)--(0.668,1.654);
\gpcolor{rgb color={0.800,0.800,0.800}}
\draw[gp path] (0.758,1.945)--(2.240,1.945);
\gpcolor{color=gp lt color border}
\draw[gp path] (0.758,1.945)--(0.713,1.945);
\gpcolor{rgb color={0.800,0.800,0.800}}
\draw[gp path] (0.758,2.114)--(2.240,2.114);
\gpcolor{color=gp lt color border}
\draw[gp path] (0.758,2.114)--(0.713,2.114);
\gpcolor{rgb color={0.800,0.800,0.800}}
\draw[gp path] (0.758,2.235)--(2.240,2.235);
\gpcolor{color=gp lt color border}
\draw[gp path] (0.758,2.235)--(0.713,2.235);
\gpcolor{rgb color={0.800,0.800,0.800}}
\draw[gp path] (0.758,2.328)--(2.240,2.328);
\gpcolor{color=gp lt color border}
\draw[gp path] (0.758,2.328)--(0.713,2.328);
\gpcolor{rgb color={0.800,0.800,0.800}}
\draw[gp path] (0.758,2.405)--(2.240,2.405);
\gpcolor{color=gp lt color border}
\draw[gp path] (0.758,2.405)--(0.713,2.405);
\gpcolor{rgb color={0.800,0.800,0.800}}
\draw[gp path] (0.758,2.469)--(2.240,2.469);
\gpcolor{color=gp lt color border}
\draw[gp path] (0.758,2.469)--(0.713,2.469);
\gpcolor{rgb color={0.800,0.800,0.800}}
\draw[gp path] (0.758,2.525)--(2.240,2.525);
\gpcolor{color=gp lt color border}
\draw[gp path] (0.758,2.525)--(0.713,2.525);
\gpcolor{rgb color={0.800,0.800,0.800}}
\draw[gp path] (0.758,2.575)--(2.240,2.575);
\gpcolor{color=gp lt color border}
\draw[gp path] (0.758,2.575)--(0.713,2.575);
\gpcolor{rgb color={0.800,0.800,0.800}}
\draw[gp path] (0.758,2.619)--(2.240,2.619);
\gpcolor{color=gp lt color border}
\draw[gp path] (0.758,2.619)--(0.668,2.619);
\gpcolor{rgb color={0.800,0.800,0.800}}
\draw[gp path] (0.758,2.909)--(2.240,2.909);
\gpcolor{color=gp lt color border}
\draw[gp path] (0.758,2.909)--(0.713,2.909);
\node[gp node center] at (0.960,0.582) {1};
\node[gp node center] at (1.230,0.582) {2};
\node[gp node center] at (1.499,0.582) {4};
\node[gp node center] at (1.768,0.582) {8};
\node[gp node center] at (2.038,0.582) {16};
\draw[gp path] (0.758,2.909)--(0.758,0.690)--(2.240,0.690)--(2.240,2.909)--cycle;
\node[gp node center] at (1.546,0.270) {\relsize{1}{Concurrency}};
\node[gp node center,rotate=-270] at (0.237,1.778) {\relsize{1}{Latency (ms)}};
\gpfill{rgb color={0.933,0.933,0.933}} (0.866,1.135)--(1.054,1.135)--(1.054,1.136)--(0.866,1.136)--cycle;
\draw[gp path] (0.866,1.135)--(1.054,1.135)--(1.054,1.136)--(0.866,1.136)--cycle;
\draw[gp path] (0.960,1.136)--(0.960,1.147);
\draw[gp path] (0.866,1.147)--(1.054,1.147);
\draw[gp path] (0.866,1.135)--(1.054,1.135);
\gpfill{rgb color={0.933,0.933,0.933}} (1.135,1.217)--(1.325,1.217)--(1.325,1.338)--(1.135,1.338)--cycle;
\draw[gp path] (1.135,1.217)--(1.325,1.217)--(1.325,1.338)--(1.135,1.338)--cycle;
\draw[gp path] (1.230,1.175)--(1.230,1.217);
\draw[gp path] (1.230,1.338)--(1.230,1.694);
\draw[gp path] (1.135,1.694)--(1.325,1.694);
\draw[gp path] (1.135,1.175)--(1.325,1.175);
\gpfill{rgb color={0.933,0.933,0.933}} (1.405,1.279)--(1.593,1.279)--(1.593,1.412)--(1.405,1.412)--cycle;
\draw[gp path] (1.405,1.279)--(1.593,1.279)--(1.593,1.412)--(1.405,1.412)--cycle;
\draw[gp path] (1.499,1.207)--(1.499,1.279);
\draw[gp path] (1.499,1.412)--(1.499,2.221);
\draw[gp path] (1.405,2.221)--(1.593,2.221);
\draw[gp path] (1.405,1.207)--(1.593,1.207);
\gpfill{rgb color={0.933,0.933,0.933}} (1.674,1.166)--(1.864,1.166)--(1.864,1.452)--(1.674,1.452)--cycle;
\draw[gp path] (1.674,1.166)--(1.864,1.166)--(1.864,1.452)--(1.674,1.452)--cycle;
\draw[gp path] (1.769,1.156)--(1.769,1.166);
\draw[gp path] (1.769,1.452)--(1.769,2.593);
\draw[gp path] (1.674,2.593)--(1.864,2.593);
\draw[gp path] (1.674,1.156)--(1.864,1.156);
\gpfill{rgb color={0.933,0.933,0.933}} (1.944,1.156)--(2.132,1.156)--(2.132,1.546)--(1.944,1.546)--cycle;
\draw[gp path] (1.944,1.156)--(2.132,1.156)--(2.132,1.546)--(1.944,1.546)--cycle;
\draw[gp path] (2.038,1.145)--(2.038,1.156);
\draw[gp path] (2.038,1.546)--(2.038,2.832);
\draw[gp path] (1.944,2.832)--(2.132,2.832);
\draw[gp path] (1.944,1.145)--(2.132,1.145);
\gpfill{color=gp lt color border} (0.866,1.136)--(1.054,1.136)--(1.054,1.136)--cycle;
\draw[gp path] (0.866,1.136)--(1.054,1.136)--cycle;
\gpfill{color=gp lt color border} (1.135,1.325)--(1.325,1.325)--(1.325,1.325)--cycle;
\draw[gp path] (1.135,1.325)--(1.325,1.325)--cycle;
\gpfill{color=gp lt color border} (1.405,1.388)--(1.593,1.388)--(1.593,1.388)--cycle;
\draw[gp path] (1.405,1.388)--(1.593,1.388)--cycle;
\gpfill{color=gp lt color border} (1.674,1.395)--(1.864,1.395)--(1.864,1.395)--cycle;
\draw[gp path] (1.674,1.395)--(1.864,1.395)--cycle;
\gpfill{color=gp lt color border} (1.944,1.480)--(2.132,1.480)--(2.132,1.480)--cycle;
\draw[gp path] (1.944,1.480)--(2.132,1.480)--cycle;
\draw[gp path] (0.758,2.909)--(0.758,0.690)--(2.240,0.690)--(2.240,2.909)--cycle;
\gpdefrectangularnode{gp plot 1}{\pgfpoint{0.758cm}{0.690cm}}{\pgfpoint{2.240cm}{2.909cm}}
\end{tikzpicture}%
%
\caption{Inference throughput and latency. \phantom{(b)}{Whiskers show min and max.}}%
\label{fig:dnninferencethroughput}%
\end{subfigure}%
\caption{\textbf{Inference is predictable in isolation (left).} Running inferences concurrently gains up to 25\% throughput (middle), at a cost of substantially increased latency variability (right) {due to interleaved GPU and OS executions.}}%
\label{fig:dnninference}
\end{figure}

We compiled ResNet50v2~\cite{zhang2020resnest} with TVM 0.7~\cite{chen2018tvm} and executed 11 million inferences in isolation on a state-of-the-art NVIDIA Tesla v100 GPU {using random inputs and batch size 1.}
We measured the latencies of each inference and show the median and high-percentile latencies in \autoref{fig:dnninferencetail}.
The 99.99$^\text{th}$ percentile latency was within 0.03\% of the median latency.

If DNN execution times can be measured and then accurately predicted for future inferences on that model, the next question is whether a distributed model serving system can preserve the predictable responsiveness of the core inference execution.

\section{Predictable Performance}
\label{sec:principles}

To build a responsive system through principled design, we further study the factors that can cause 
or amplify performance variability.
Importantly, components at any level of the modern system stack can contribute to variable request latency,
whether at the application layer, in the operating system, or even in the hardware~\cite{li2014tales}.
Network effects and workload fluctuations add two more sources
of unpredictability to distributed systems.

\fakepara{The whole is more than the sum of its parts.} The overall system performance variability is primarily governed by how the system is assembled from  its constituent components.
We can handle variable latency of a software component in several ways. 
First, we can ignore the problem and allow the volatility to propagate to later requests or percolate to other components of the system.
Even performance-conscious code that is optimized to improve throughput or average latency does not fix tail latency~\cite{delimitrou2018amdahl}.
An example of this contagiousness of unpredictability, known as the ``straggler'' problem in data analytics frameworks~\cite{ousterhout2015making,ananthanarayanan2013effective}, is when
a worker executes a request that takes unusually long and the other requests that were enqueued on the worker in the meantime then incur the extra delay from the unexpected wait-time.
Ignoring the variability can further compound the problem across the system, such as when the request handler itself has variable latency~\cite{vigfusson2010dr}.

Second, we can mitigate the volatility by ensuring all requests match the worst-case latency,
thus exchanging lower resource utilization for predictability---often a steep price when worst-case latency
is significantly higher than the median.

Third, we can minimize variability by expending more resources, again in trade for lower utilization.
Some networked systems, for instance, 
are designed to submit the same job to multiple workers in parallel and then to cancel unneeded jobs upon successfully receiving a result from the fastest worker~\cite{dean2013tail}. 

Fourth, upon detecting an unusual delay, we can notify a feedback mechanism to adjust the environment to lower the  impact on future requests. 
Such ``best-effort'' methods are typically reactive and aimed at longer-term effects, such as by temporarily adding more resources (auto-scaling~\cite{gandhi2012autoscale}), throttling requests, or balancing load. %

\fakepara{Consolidating choice.}
We take a fundamentally different approach: \emph{designing a predictable system from the bottom up.}
Our strategy is to {restrict the choices available to lower system layers as much as possible}---a 
philosophy based on our observation that when executing an essentially predictable task, performance variability only arose when a lower layer in the system was given choices regarding how to execute its task.
Examples from all layers of the systems stack abound, including:
\begin{compactitem}
  \item \textbf{Hardware level:} when a GPU is passed multiple CUDA kernels to execute in parallel, the GPU has the choice of how to allocate resources, including execution units and memory bandwidth, between kernels.
    The GPU makes these choices based on its internal state and undocumented, proprietary policies.
  \item \textbf{OS level}: when we create multiple threads that the operating system can execute on the same core, the OS has the choice of what threads to execute when, based on internal scheduling policies and state.
  \item \textbf{Application level}: when the worker processes of a distributed application each manage their own cache independently, the workers have the choice of what to cache and for how long, leading to unpredictable hit rates and latency variability~\cite{huang2013analysis}; similarly, when worker processes implement their own thread pools and queuing policies, they have the choice of which requests to execute first, leading to unpredictable queuing times.
\end{compactitem}
\autoref{fig:dnninferencethroughput} illustrates this: a standard design for building a worker would use thread pools serving inference requests in parallel to saturate the GPU.
While concurrent threads indeed increase inference throughput by up to 25\%, the factors above cause tail latency to increase by $100\times$.

Our {approach} is to \emph{consolidate choices} in the upper layers: once a layer implements choices for lower layers based on internal state, it forces the lower layer to follow a narrow path of possible executions, causing the performance of the resulting layer to be nearly deterministic.
The upper layer can then {sufficiently} predict the performance of the lower layers and reason with foresight about  resource utilization and the anticipated execution times for all requests.
The price of this strategy, however, is a tighter coupling of components and a less modular architecture.

\fakepara{Imperfect predictability.}
{Notably, we can consolidate choice without requiring \emph{perfect} predictability. Real systems will retain some unpredictable components, such as managing CPU caches or workload shifts, even after consolidating choices in its upper layers. Instead, the chief goal of concentrating these choices is to make predictable executions the common case.  This frees us from implementing best-effort mechanisms to tolerate the occasional, rare instance of unpredictability; instead unpredictability can be directly treated as an error.}

\section{Design} %

\newcommand{\modelicons}{%
\tikzsetnextfilename{modelicons}%
\begin{tikzpicture}
\begin{scope}[scale=0.65,every node/.append style={inner sep=0,outer sep=0,transform shape}]
	\tikzstyle{model}=[draw=black!50]
	\definecolor{myturquoise}{HTML}{B3CEE1}
	\definecolor{myorange}{HTML}{FDCDAC}
	\definecolor{myblue}{HTML}{CBD5E8}
	\definecolor{mypink}{HTML}{F4CAE4}
	\definecolor{mygreen}{HTML}{E6F5C9}
	\definecolor{myyellow}{HTML}{FFF2AC}
	\definecolor{mybrown}{HTML}{F1E2CC}
	\definecolor{mygray}{HTML}{CCCCCC}
	\definecolor{mydarkgray}{HTML}{666666}
	\tikzstyle{model1}=[model, star,star points=5,star point ratio=0.4, minimum width=4, minimum height=4, rotate=180, fill=myyellow]
	\tikzstyle{model2}=[model, regular polygon,regular polygon sides=7, minimum width=10, minimum height=10, fill=mybrown]
	\tikzstyle{model3}=[model, regular polygon,regular polygon sides=4, minimum width=11, minimum height=11, fill=myturquoise]
	\tikzstyle{model4}=[model, star,star points=7,star point ratio=0.6, minimum width=7, minimum height=7, fill=mygreen]
	\tikzstyle{model5}=[model, diamond, minimum width=10, minimum height=10, fill=mygray]
	\tikzstyle{model6}=[model, regular polygon,regular polygon sides=5, minimum width=10, minimum height=10, fill=mypink]
	\tikzstyle{model7}=[model, star,star points=8,star point ratio=0.3, minimum width=3, minimum height=3, rotate=180, fill=myblue]
	\tikzstyle{model8}=[model, regular polygon,regular polygon sides=3, minimum width=11, minimum height=11, fill=myorange, yshift=-1]
	\node [model3] at (0.33,-0.15) {};
	\node [model8] at (0.18,0) {};
	\node [model1] at (0,-0.1) {};	
\end{scope}
\end{tikzpicture}}

\newcommand{\modelstar}{%
\tikzsetnextfilename{modelstar}%
\begin{tikzpicture}
\begin{scope}[scale=0.65,every node/.append style={inner sep=0,outer sep=0,transform shape}]
	\tikzstyle{model}=[draw=black!50]
	\definecolor{myyellow}{HTML}{FFF2AC}
	\tikzstyle{model1}=[model, star,star points=5,star point ratio=0.4, minimum width=4, minimum height=4, rotate=180, fill=myyellow]
	\node [model1] at (0,0) {};	
\end{scope}
\end{tikzpicture}\xspace}

\newcommand{\modeltriangle}{%
\tikzsetnextfilename{modeltriangle}%
\begin{tikzpicture}
\begin{scope}[scale=0.65,every node/.append style={inner sep=0,outer sep=0,transform shape}]
	\tikzstyle{model}=[draw=black!50]
	\definecolor{myorange}{HTML}{FDCDAC}
	\tikzstyle{model8}=[model, regular polygon,regular polygon sides=3, minimum width=11, minimum height=11, fill=myorange, yshift=-1]
	\node [model8] at (0,0) {};	
\end{scope}
\end{tikzpicture}\xspace}

\begin{figure}[t]
\centering%
\input{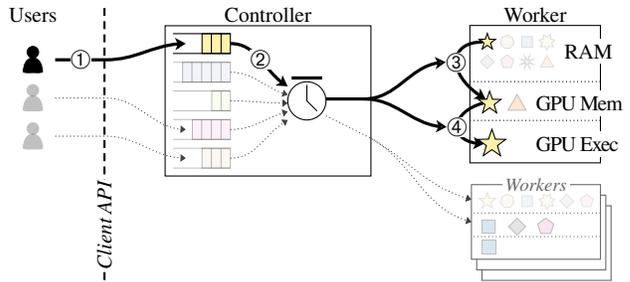}%
\caption{{Clockwork comprises multiple Workers and a centralized Controller.  Models (\protect\modelicons) reside on Workers; inference requests are queued and scheduled centrally on Clockwork's Controller.  See \autoref{sec:overview} for a detailed description.}}
\label{fig:architecture}
\end{figure}

By recursively {restricting} choice from lower layers, we converge on a  design where {the most} performance-critical execution choices are made in the topmost layer.
In the context of a model serving service, this process converges to an architecture, which we call \sysname{}, with a centralized controller and workers with predictable performance.

\subsection{Overview}
\label{sec:overview}

\fakepara{Architecture.}
{\autoref{fig:architecture} illustrates Clockwork's architecture.  Users submit inference requests (\circone) which are queued centrally on Clockwork's controller.  Each worker has a set of DNN models (\modelicons) loaded into RAM and maintains exclusive control over one or more GPUs.  The centralized scheduler has a global view of system state, including all workers, and decides when to execute each request (\circtwo).  To execute a request, the scheduler explicitly decides when to load models into GPU memory (\circthree) and when to execute requests on the GPU (\circfour).  At any time, the scheduler makes accurate, high-quality caching, scheduling, and load balancing decisions.  The controller can perform these actions proactively because execution on workers is highly predictable.  
The controller transmits continual scheduling information to the workers that, by design, will execute schedules exactly as directed.}

\begin{figure}[t]%
\input{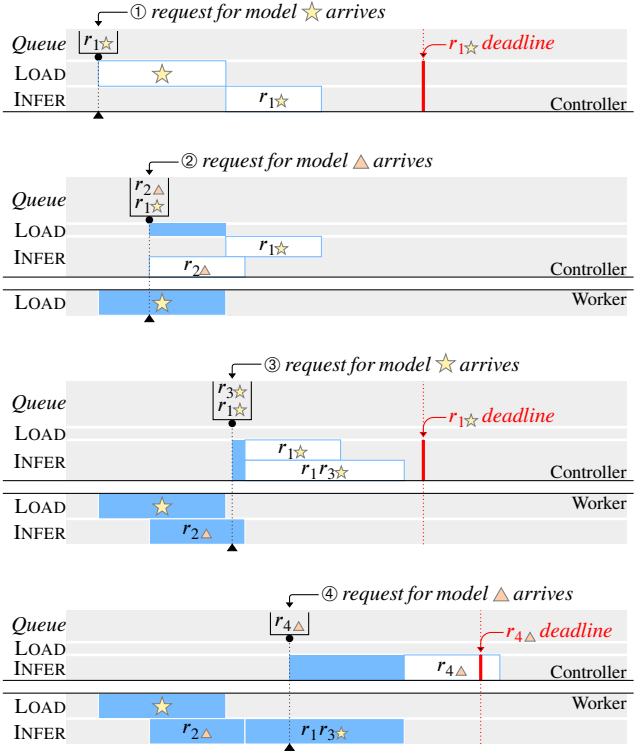}%
\caption{Timeline of four illustrative inference requests.}%
\label{fig:timeline}%
\end{figure}

\fakepara{Illustrative example.}
To elucidate the \sysname{} architectural components with more detail, including the choices that were consigned to the controller, consider the key steps for serving the inference requests illustrated in~\autoref{fig:timeline}.

\circone Upon receiving an inference request $r_1$ for model \modelstar, the controller is aware that a target worker has yet to copy the model weights from RAM into GPU memory.  It estimates the time required to load the model weights (\loadweights), plus the time to subsequently execute the inference (\infer), and concludes that the request will complete within its specified SLO.  The controller instructs the worker to copy the model weights to GPU memory via a \loadweights action.  Since the controller is aware of all timings, it does not yet need to submit the subsequent \infer action until the \loadweights has completed.

\circtwo While \modelstar is loading, a request $r_2$ for model \modeltriangle arrives.  The controller is aware that, unlike \modelstar, \modeltriangle is already loaded into GPU memory.  The controller can choose to either \infer $r_2$ immediately, or wait for \modelstar to complete loading then \infer $r_1$.  Since the worker would be otherwise idle, the controller instructs the worker to execute the inference for $r_2$ immediately via an \infer action.

\circthree Clockwork workers only execute one \infer action and one \loadweights action at a time, so the controller can wait until $r_2$  has nearly completed before submitting an \infer action for $r_1$.  In the meantime, another request $r_3$ for model \modelstar arrives.  This gives the controller a choice between \infer for $r_1$ by itself, or to \emph{batch} $r_1$ and $r_3$.  Batched execution is more efficient, but takes longer.  In this case a batched \infer action will still complete before $r_1$'s deadline, so the controller instructs the worker to batch the inferences for $r_1$ and $r_3$.

\circfour While $r_1$ and $r_3$ execute, a request $r_4$ for \modeltriangle arrives with a tight SLO.  The controller is aware that $r_4$ will miss its deadline, even if it executes immediately after the worker becomes free.  The controller does not proceed to schedule an \infer action, and cancels the request before performing any fruitless work.

Each step of the above execution is fast, \eg for ResNet50, \loadweights and \infer take approximately 8\,ms and 3\,ms respectively.  \autoref{table:models} outlines representative measurements for 8 of the 61 models used for Clockwork experiments.

\begin{table*}
\centering%
\newcommand{\thbc}[1]{\multicolumn{1}{c|}{\textbf{#1}}}%
\footnotesize%
\setlength\extrarowheight{1.5pt}%
\setlength\dashlinedash{\arrayrulewidth}%
\setlength\dashlinegap{1.5pt}%
\setlength\tabcolsep{5pt}%
\setlength\arrayrulewidth{0.3pt}%
\begin{tabular}{l|l|r|r|r|r|r|r|r|r|r}
  \multirow{2}{*}{\textbf{Model Family}} &
  \multirow{2}{*}{\textbf{Model}} &
  \multicolumn{2}{c|}{\textbf{IO Size (kB)}} &
  \multicolumn{2}{c|}{\textbf{Weights}} &
  \multicolumn{5}{c}{\textbf{GPU Execution Latency (ms)}} \\
  \cline{3-11}
  & & \thbc{Input} & \thbc{Output} & \thbc{Size (MB)} & \thbc{Transfer (ms)} &
  \thbc{B1} & \thbc{B2} & \thbc{B4} & \thbc{B8} & \multicolumn{1}{c}{\textbf{B16}} \\
\hline

\multirow{1}{*}{DenseNet~\cite{huang2017densely}}
& densenet169 & 602 & 4 & 56.5 & 4.50 & 5.18 & 6.29 & 8.57 & 12.82 & 21.85 \\
\cdashline{1-11}
\multirow{1}{*}{Inception v3~\cite{szegedy2016rethinking}}
& inceptionv3 & 1073 & 4 & 95.3 & 7.77 & 4.46 & 6.85 & 10.99 & 16.45 & 26.17 \\
\cdashline{1-11}
\multirow{1}{*}{Mobile Pose~\cite{xiao2018simple}}
& mobile\_pose\_mobilenetv3 & 590 & 209 & 19.0 & 1.55 & 1.29 & 1.92 & 3.13 & 5.71 & 11.62 \\
\cdashline{1-11}
\multirow{3}{*}{ResNet~\cite{he2016deep}}
& resnet18 & 602 & 4 & 46.7 & 3.81 & 1.27 & 1.86 & 2.73 & 4.06 & 7.02\\*
& resnet50 & 602 & 4 & 102.3 & 8.33 & 2.61 & 3.78 & 5.61 & 9.13 & 15.67 \\*
& resnet152 & 602 & 4 & 240.9 & 19.58 & 7.71 & 11.14 & 16.21 & 26.48 & 44.60 \\*
\end{tabular}%
\caption{{Measurements of a representative subset of the 61 models used for \sysname experiments.  Pre-trained models were sourced from the ONNX Model Zoo~\cite{onnxmodelzoo} and the GluonCV Model Zoo~\cite{guo2020gluoncv}, and optimized for NVIDIA Tesla v100 GPUs using TVM v0.7~\cite{chen2018tvm}.}}
\label{table:models}
\end{table*}

\subsection{Consolidating Choice}
{Our design consolidates choice in three main ways.}
First, changes in the worker's state, for instance evicting a DNN from GPU memory, 
can influence the performance for future requests in a way that makes performance estimation complex.
We therefore require that no worker operation should have implicit performance side-effects on any future operation. 
Second, we must ensure that a predictable component either delegates
scheduling decisions that may impact performance to the centralized controller, or otherwise
makes schedules deterministic. 
Third, when a predictable component is unable to execute a schedule as instructed, it is
treated as an error to enable workers to get back on schedule. 
Workers do not attempt best-effort remediation, so as to avoid
a cascade of mispredictions.

We enforce these three properties in \sysname{} through an action command abstraction 
between the controller and workers that, in lieu of traditional RPC calls, 
either communicates a change in a worker's state or a task for a worker to execute.
Each action the controller issues to a worker, such as \loadweights{} and \infer{}, 
has predicted execution time and a designated execution window.
These are derived using the known state of the worker, previously submitted actions,
and known transitions in controller-maintained worker state.

\subsection{Challenges for Predictable Inference}

To consolidate choice we must 
first identify where performance-critical choices arise in system components.
We have established that DNN inference itself on a GPU has deterministic performance; we next study the challenges in extending this result to a full-fledged inference system.

\newcommand{\challmemman}{\textbf{C1}\xspace}
\fakepara{Managed memory and caches can be unpredictable (\challmemman).}
RAM and GPU memory on a worker constitute state that impacts the performance of future requests. 
Additionally, some memory allocators exhibit variable timing for allocation and deallocation requests due to internal trade-offs between memory fragmentation and amortized performance.
Memory that is used as a cache specifically introduces performance variability between
cache hits and misses, with an internal cache replacement policy influencing performance
of future items.
To maintain predictability, we must instead consolidate choice by managing cache admission and eviction for each worker at the central controller.  Fortunately, caching of DNN weights is coarse-grained and per-model.

\newcommand{\challhwsched}{\textbf{C2}\xspace}
\fakepara{Hardware interactions can be unpredictable (\challhwsched).}
Many system resources are implicitly administered by hardware schedulers that operate at very fine time-scales and produce different schedules under even minute shifts in the arrival times of other requests. 
The volatility of timing coupled with proprietary and un-documented scheduling policies make it onerous to accurately predict completion times for concurrent requests.
The remedy for non-determinism is to strip away the ability for schedulers to reorder requests by forcing only a single request to be executed at a time, at the cost of spending greater effort on keeping the resource fully utilized. Mercifully, one-at-a-time execution of DNN inferences on GPUs has closely comparable throughput to concurrent execution (\autoref{fig:dnninferencethroughput}) and many classes of DNNs (\eg convolutional neural networks) can saturate GPUs with small batch sizes.

\newcommand{\challexternal}{\textbf{C3}\xspace}
\fakepara{External factors can trigger performance variance (\challexternal).}
Even after systematically removing the key internal sources of unpredictability by consolidating choice, there will always remain external sources outside of the controller's purview. These include performance interference through shared network bottlenecks, thermal throttling of CPUs and GPUs, and others.
The only option is to minimize their effects by building sufficient tolerance into the system.

\subsection{Predictable DNN Worker}

At a high-level, \sysname{} workers maintain DNNs in memory and execute inference requests on one or more GPUs.
The workers interface with the controller to receive actions.

\fakepara{Memory management.}
Model weights must be present in GPU memory to execute an inference.
However, GPU memory capacity is small (\raisebox{.2ex}{$\mathsmaller\leq$}32GB) relative to host memory (\raisebox{.2ex}{$\mathsmaller\leq$}4TB), and host-to-GPU memory transfers 
 (\raisebox{.2ex}{$\mathsmaller\approx$}8.3ms for ResNet50) typically take longer than running the DNN inference on the GPU (\raisebox{.2ex}{$\mathsmaller\approx$}2.9 ms).
Consequently, \sysname{} treats GPU memory as a cache, letting commonly or recently used models avoid expensive loads. 
To overcome \challmemman{}, workers explicitly expose \loadweights and \unloadweights actions to the controller for copying models to and removing models from worker's GPU memory with deterministic latency.
These actions also update the state that the controller tracks for the worker.

\fakepara{Inference execution.}
The controller only sends an \infer action when a model is present in GPU memory or a \loadweights action will momentarily complete.
The worker
internally divides \infer actions into three steps.
First, \copyinput transfers the input vector from host to GPU memory. 
Next, \exec performs the actual heavy-weight DNN GPU calculations, which dominate the total inference time.
Finally, \copyoutput transfers the resulting output vector from the GPU back to host memory.
These steps may coincide: the previous request's outputs can be copied at the same time as the current request's input is being transferred.
However, multiple concurrent \exec{} calls cause the GPU hardware scheduler to behave unpredictably (\challhwsched). 
Fortunately, a DNN inference call by itself can efficiently utilize the GPU while also restricting the hardware scheduler to a single, predictable option (\autoref{fig:dnninferencethroughput}).
\sysname{} workers therefore run a single \exec{} at a time, 
a design choice that reduces performance variability by two orders of magnitude while only minimally decreasing inference throughput (\autoref{fig:dnninferencethroughput}).

\fakepara{Interface with the controller.}
\sysname workers receive \loadweights{}, \unloadweights{}, and \infer{} actions from the controller
with detailed timing expectations attached:

{
\rowcolors{2}{blue!15}{blue!5}
\noindent
\resizebox{\columnwidth}{!}{
\begin{tabular}{ll}
\rowcolor{blue!15}
\texttt{type} & \infer, \loadweights, or \unloadweights\\
\texttt{earliest} & the time when this action may begin executing \\ 
\texttt{latest} & when this action will be rejected
\end{tabular}
}
}

Rather than executing actions in a work-conserving, best-effort manner, workers strictly follow the schedule of actions imposed by the controller.
The controller communicates two timestamps with every action, \texttt{earliest} and \texttt{latest}, to designate a time interval during which the worker may begin executing the action.
Actions that cannot start within the prescribed window are cancelled and never executed.
This allows workers to quickly get back on schedule after an individual action is delayed unexpectedly (\challexternal{}) by skipping one or more actions, minimizing the impact of the delay on other actions.
Workers communicate the result of each action back to the controller, including whether the command was successful and the measured execution time.

\subsection{Central Controller}
All decision-making in \sysname{} occurs in the central controller.
The controller receives inference requests from users and decides worker actions
while striving to meet SLOs. 

\paragraph{Modeling worker performance.}
The controller maintains a per-worker, per-model performance profile comprising processing time measurements of recent requests; profiles are updated continuously to tolerate shifts due to external factors (\challexternal{}).
The controller also tracks the outstanding actions and memory state at every worker. 
Since actions have inherently deterministic latency by design, the controller can deduce the earliest time that a worker could begin executing a new action (queuing time).

\paragraph{Action scheduler.}
The Clockwork controller proactively manages action schedules for workers.  It utilizes a global view of system requests, up-to-date worker performance profiles, and accurate predictions for when outstanding actions will complete.
The controller attempts to pack worker schedules tightly by
making narrow, realistic estimates for the \texttt{earliest} and \texttt{latest} time interval.
The interval width balances a trade-off between \sysname{} SLO fulfillment and system goodput.
On one hand, making the interval too narrow increases the risk of an action not being executed by a worker 
because it could not be completed in time (\challexternal{}), potentially triggering an SLO violation. 
On the other hand, underestimating the window length can create periods of inactivity and decrease worker utilization, thus affecting \sysname{} goodput.

The scheduler lazily decides which worker should execute the inference.
The controller only submits a minimal amount of work to keep workers utilized; it is in no hurry to commit because it can accurately predict action timings.
Delaying choices on the controller improves schedules by providing more options, permitting the \sysname controller to re-order and \emph{batch} inference requests to the same model, significantly improving resource efficiency and throughput.

In our  design, any worker can process any request since they all store every model in host memory; however, workers have different sets of models loaded into their GPU memory.
A worker that executes only cold inferences must transfer weights for each model from host memory to the GPU and may saturate the available PCIe bandwidth, whereas a worker that executes only hot inferences may be bottlenecked by the GPU.
The \sysname{} scheduler balances load by mixing and matching hot and cold inferences among all workers.

\section{Implementation}
\label{sec:implementation}

\sysname{}'s implementation, comprising 26KLOC of C\texttt{++}, contains various decisions that enable \sysname{} to consolidate choice on its controller.

\subsection{Models}

\fakepara{Predictable model execution.}  
Prior model serving systems such as Clipper~\cite{crankshaw2017clipper} and INFaaS~\cite{romero2019infaas} act as orchestration layers atop existing model execution frameworks such as TensorFlow~\cite{abadi2016tensorflow} and TensorRT~\cite{tensorrt}.
This decoupling makes it difficult to consolidate choice, since the model execution frameworks encapsulate scheduling and memory management decisions that we wish to make with \sysname{}.  Instead, \sysname{} implements its own model runtime, reusing key components of the TVM optimizing compiler~\cite{chen2018tvm}.  
\sysname{}'s model runtime enables fine-grained control over each stage of a model's execution.  
For models provided to \sysname{} (\eg in ONNX form), we compile a binary representation using TVM and postprocess the model to produce the following:
\begin{compactitem}
\item \textbf{Weights:} A model's weights are a binary blob (10s to 100s of MB (\cf~\autoref{table:models}).  
\item \textbf{Kernels:} The CUDA kernels that execute a model (10s to 100s of kB).  These are not provided by the user; they are derived from the abstract model definition, and kernels from different users can safely execute within the same process.  \sysname{} uses the kernels compiled by TVM.  \sysname{} compiles kernels for multiple configurable batch sizes; by default 1,2,4,8, and 16.  Kernels for different batch sizes can use the same weights without modification.
\item \textbf{Memory metadata:} At runtime, models do not directly allocate memory; instead, \sysname{} will pre-allocate and manage all GPU memory and pass pointers as arguments to function calls.  The memory requirements for a model are static, and \sysname{} precalculates the required workspace memory and offsets required for each kernel.
\item \textbf{Profiling data:} \sysname{} runs a brief profiling step to produce a seed estimate for model execution times.
\end{compactitem}

\fakepara{Model loading.}
Models are stored in an efficient serialized form on disk.  \sysname{} workers pre-load models from disk into main memory on worker startup.  For the worker machines used in our evaluation, 768GB RAM can support thousands of models (\cf~\autoref{sec:azure_prediction_errors}).  Once a model is in main memory, \sysname{} extracts and links the CUDA modules needed for its execution.  To improve predictability, \sysname{} disables JIT compilation and the caching of CUDA kernels.  

\subsection{DNN Workers}

Each machine runs one worker process that receives and executes actions from \sysname{}'s controller.  We do not run \sysname{} in a container or VM to avoid the performance interference such sharing can impose.

\fakepara{Managing model weights in memory.}
\sysname{} pre-allocates all GPU memory and divides it into three categories:
\begin{compactitem}
\item \textbf{Workspace:} Models require a variable amount of GPU memory for intermediate results.  This memory is transient and only needed during execution; once an output has been produced, it is no longer needed.  \sysname{} only executes models one-at-a-time, so it allocates 512MB workspace memory.
\item \textbf{IOCache:} Although \sysname{} only executes models one-at-a-time, \sysname{} asynchronously copies inputs to the GPU prior to execution, and outputs to host memory after execution.  \sysname{} allocates 512MB device memory for temporary storage of inputs and outputs before and after execution.
\item \textbf{PageCache:} The remaining device memory is used for storing model weights, divided into 16MB \emph{pages}.
Multiple tensors can occupy the same 16MB page and the mapping of tensors to pages is determined statically at model-compile time. At runtime, page pointers are passed as kernel arguments and tensors are read from pre-defined offsets.
\end{compactitem}

\sysname{}'s PageCache has several advantages.  First, avoiding repeated memory allocation calls leads to more predictable executions, since memory allocation can be an unpredictable source of overheads (\challmemman{}).  Second, paging \emph{simplifies choice}: external memory fragmentation issues are eliminated, and the controller need only track the number of total free pages to completely capture the worker's memory state.  Paging slightly increases memory utilization; however, model memory requirements are static and known ahead of time, and can be bucketed on to pages to reduce internal fragmentation.  Paging does not affect the latency of memory transfers.

\fakepara{Actions.}
To orchestrate workers, the controller uses the previously described \emph{action} abstraction.
Actions contain a unique \texttt{id} and an action-dependent \texttt{payload} (\eg \infer inputs).  Each worker runs a dedicated \emph{executor} for each action type and each worker-GPU.  An executor runs a thread that dequeues actions chronologically by \earliest timestamp, and waits until \earliest is reached before proceeding with an action.   Executors reject actions whose \latest timestamp has passed.  To reduce interference between threads and other processes, each executor is pinned to a dedicated core and runs at real-time priority.  Both \infer and \loadweights execute asynchronous work in their own CUDA streams.  Each executor is bottlenecked by a different resource (\eg GPU execution and PCIe transfers) %
and can run concurrently with negligible interference.

\fakepara{Results.}
A network thread maintains a persistent connection with the controller for receiving actions and sending results.  A result comprises the following:

{
\rowcolors{2}{blue!15}{blue!5}
\noindent
\resizebox{\columnwidth}{!}{
\begin{tabular}{ll}
\rowcolor{blue!15}
\texttt{status} & success or an error code \\
\texttt{timing} & start and end times, and on-device execution \\ 
\rowcolor{blue!5}
& duration for any asynchronous work
\end{tabular}
}
}

\loadweights actions acquire pages from the PageCache, then copy weights to those pages. If no pages are available then \loadweights aborts.
The controller explicitly frees pages with \unloadweights{}; this only updates in-memory metadata and always succeeds.

\infer actions comprise \copyinput, \exec and \copyoutput, each of which have dedicated executors.  \copyinput executes immediately on receipt of \infer; it acquires IO memory from the IOCache then copies inputs.  \exec inherits the \infer action's \earliest and \latest timestamps; it checks weights and inputs are present then executes kernels on the GPU, using Workspace for intermediate calculations.  \copyoutput immediately copies outputs back to main memory then releases the IO memory.  To simplify controller decision making, \copyinput and \copyoutput are not exposed as actions since they are orders of magnitude faster than \exec and \loadweights (10s of microseconds) for our workloads.  \sysname{}'s memory management allows for back-to-back \infer actions for the same model.

\begin{figure*}%
\tikzsetnextfilename{comparison_experiment}%
\input{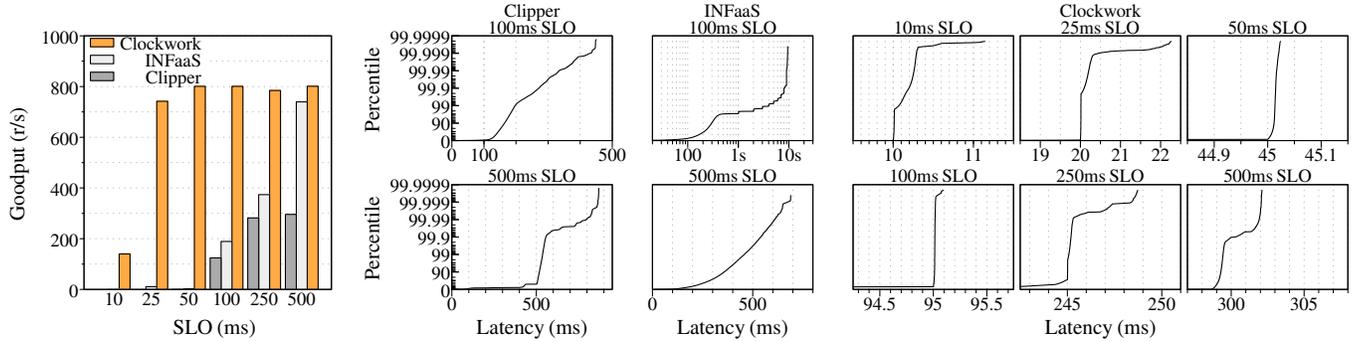}%
\vspace{-5mm}%
\caption{\textls[-5]{Goodput and latency measurements for Clipper, INFaaS, and \sysname.  We deploy 15 instances of ResNet50 on 1 worker; each model submits 16 concurrent requests in a closed loop.  (Left) Request goodput.  Goodput only counts requests that succeed within the SLO.  (Right) Request latency CDFs across all requests (including those rejected due to missed deadlines).  Latency CDFs are scaled to highlight tail latency.}}%
\label{fig:simple_experiment}%
\end{figure*}

\subsection{Central Controller}

On startup, Clockwork's controller establishes persistent connections to all workers and exchanges metadata about the size of each worker's PageCache, the models present on each worker, and their initial pre-profiled execution times.  The core duty of the controller is to satisfy requests received from clients by submitting actions to workers.  This decision making is encapsulated in the \emph{Scheduler} interface:

{
\rowcolors{2}{blue!15}{blue!5}
\noindent
\resizebox{\columnwidth}{!}{
\begin{tabular}{ll}
\rowcolor{blue!15}
\texttt{onRequest} &  client request received, specifying a model\\
\rowcolor{blue!15}
& ID, SLO, and providing inference inputs \\
\rowcolor{blue!5}
\texttt{onResult} &  a result is received from a worker 
\end{tabular}
}
}

A scheduler implements this interface, and can invoke \texttt{sendAction} to send an action to a worker, and \texttt{sendResponse} to respond to a client.  A separate layer of the controller implements common tasks such as networking, forwarding inputs to workers, setting timestamps, and handling timeouts.  This design concentrates all choice in a single place, and enables different scheduler implementations to be easily dropped in.

\fakepara{Managing worker state.}
The controller maintains an accurate representation of workers' execution state, which is threefold: 
\emph{memory state}, in which the scheduler tracks what models are present in the worker PageCaches and when \loadweights{} will be required; 
\emph{action profiles}, which are measurements of past 10 actions duration, stratified by model, worker, and batch size, to predict the duration of future action; and \emph{pending actions}, which tracks submitted actions and estimates when each executor will next be available.  Taken together, these enable the scheduler to accurately predict when candidate actions will complete, and avoid submitting work that cannot complete before the request's deadline.  Worker state is not a significant scalability bottleneck; action profiles require only 40 bytes for each model, worker and batch size combination.

\fakepara{Scheduling \infer.}
Upon arrival, requests are enqueued into per-model request queues.  For each \infer executor, a new action must be scheduled whenever the executor has less than 5\,ms of outstanding work.  To schedule an \infer action, a model and batch size must be selected.  The batch size can differ action-to-action, though the scheduler prioritizes larger batch sizes for efficiency.

At any point in time, a model will have zero or more queued requests.  However, not every request is suitable for every batch size.  Higher batch sizes take longer to execute, so a request close to its deadline might only be satisfiable using a small batch size.  To handle this, each model has a request queue per batch size (we term this a \emph{batch queue}).  New requests are enqueued into \emph{every} batch queue.  Requests are dropped from batch queues when they cease to be satisfiable; \eg{} a request in the batch size of 16 queue will be dropped sooner than it is dropped from the batch size of 8 queue.

To decide which model and batch size to schedule, we use \emph{strategies}.  A strategy specifies a \emph{model}, a \emph{latest} timestamp, and a \emph{batch size}. Each \infer executor has a separate strategy queue, ordered by \emph{latest}, containing only strategies for models it has loaded.  The scheduler dequeues strategies until it finds one that is \emph{valid}: \emph{latest} has not elapsed, and the batch queue for the specified batch size has sufficient requests.  If a strategy is valid, the scheduler will also speculatively increase the batch size as long as extra requests are available.

When a valid strategy is found, an \infer action is created and requests are dequeued to fill the batch.  Old strategies for this model are removed from the strategy queue, and  new strategies are then created and enqueued.  A strategy is created per batch queue; \emph{latest} is calculated by subtracting the batch execution time from the deadline of the request at the head of the queue.  Empty batch queues are skipped.

\fakepara{Scheduling \loadweights.}
Each \loadweights executor also schedules up to 5\,ms of outstanding work.  For a \loadweights executor, the scheduler selects a model by estimating each model's SLO violations given the model's current state and outstanding requests.  To do this efficiently, the scheduler maintains and incrementally updates \emph{load} and \emph{demand} statistics for models and GPUs:
\begin{compactitem}
\item $d_m$ \hspace{1mm} the total demand for each model $m$
\item $a_{m,g}$ \hspace{1mm} the demand allocation of model $m$ on GPU $g$.
\item $\ell_g=\sum_ma_{m,g}$ \hspace{1mm} the total load on each GPU $g$
\end{compactitem}
A model's total demand $d_m$ is the total estimated execution time of $m$'s outstanding requests; we update $d_m$ when requests for that model arrive and complete.  The demand allocations $a_{m,g}$ for $m$ on GPU $g$ are also updated when requests arrive and complete; they are calculated such that $\sum_{g} a_{m,g}=d_m$.  Demand allocations are 0 for GPUs where the model is not loaded.  On GPUs where the model is loaded, demand allocations are inversely proportional to the GPU's load, since overloaded GPUs will be able to execute proportionally less of the total demand.  Each GPU's total load $\ell_g$ is the sum of its allocations across all models.
With these estimates, each model's load priority is defined as \vspace{-2mm}\[ p_{m,g}=d_m - \displaystyle \sum_g a_{m,g} \cdot \frac{\text{capacity}_g}{\ell_g}.\vspace{-2mm}\]
A model's load priority estimates its unfulfilled work.  For example, a model that is not loaded on any GPUs has priority equal to its outstanding work; a model loaded on a GPU that sits mostly idle has negative priority since the GPU can serve more work than the model demands.  

\sysname{} does not attempt to converge to a perfect demand allocation each time the system's state changes. Rather, \sysname{} incrementally updates each model's demand allocation and load priority 
\textbf{(i)} when new requests arrive for that model; 
\textbf{(ii)} when an \infer is initiated for that model; 
\textbf{(iii)} when \loadweights and \unloadweights affect a model; and
\textbf{(iv)} 
when a request crosses the point where it can benefit from \loadweights before its deadline.

The scheduler selects \loadweights actions by choosing the highest priority model that is not already loaded. Notably, models with negative priority need not be loaded since their demands are already met. %
\sysname{} uses a least-recently-used (LRU) eviction policy
when selecting models to \unloadweights.

\section{Evaluation}
\label{sec:evaluation}

We next assess \sysname's ability to reliably serve DNNs under a variety of workload conditions.
We begin our experimental evaluation with simple workloads in controlled settings, before expanding to heterogeneous models and diverse workloads.
Our evaluation shows that \sysname's assumptions about predictability hold, and result in a system that can effectively meet SLOs and drastically reduce tail latency.  

\fakepara{Experimental setup.}
We deploy \sysname{} in a private cluster of 12 Dell PowerEdge R740 Servers.  Each server has 32 cores, 768\,GB RAM, and 2$\times$NVIDIA Tesla v100 GPUS with 32\,GB memory.  The servers are connected by 2$\times$10\,Gbps Ethernet on a shared network.  In all experiments, we run the controller, clients, and workers on separate machines.

\subsection{How Does \sysname{} Compare?}
\label{sec:evaluation-compare}

We begin with a comparison to two prior model serving systems, Clipper~\cite{crankshaw2017clipper} and INFaaS~\cite{romero2019infaas}.  For Clipper and \sysname{}, we provision a single cluster machine to  use 1 GPU to serve 15 separate copies of ResNet50. ResNet50 is the \emph{de facto} model used for comparison previously by these systems; we chose 15 models as this reached the memory limit of Clipper\footnote{INFaaS memory limits were reached at 64 models}.  To evaluate INFaaS, we deployed an \texttt{m5.24xlarge} and a \texttt{p3.2xlarge} EC2 instance as the master and the worker, respectively.  These are not identical experiment conditions; however, INFaaS is tightly integrated with EC2, and could not be deployed on our cluster infrastructure.  We include these results for qualitative comparison.

\fakepara{Offered load.}
For each model, we run 16 closed-loop clients\footnote{Open-loop clients yielded similar results}.  The serving systems may batch requests for the same model instance, but requests to different instances cannot be batched.
We run multiple experiments, varying the target SLO from 10\,ms to 500\,ms.  

\fakepara{Goodput.}
\autoref{fig:simple_experiment} plots the \emph{goodput} achieved by each system as the target SLO varies from 10\,ms to 500\,ms.  Goodput is the number of successful requests that completed within the target SLO; it excludes timed out requests and requests that responded after the SLO.

With a high SLO of 500\,ms, \sysname and INFaaS meet their SLOs and have comparable goodput of approximately 800\,r/s.  Clipper's goodput is substantially lower, as Clipper only treats SLOs as an average latency target, not a strict threshold, and converges to this target over time without bounding latency variability.  As SLOs tighten, goodput and tail latency deteriorate for both Clipper and INFaaS, and their goodput collapses below a 100\,ms SLO.  Like Clipper, INFaaS uses the SLO as a coarse-grained goal for reactive policies.  Consequently, only \sysname can continue serving SLOs below 100\,ms.

\autoref{fig:simple_experiment} also plots latency CDFs for Clipper, INFaaS, and \sysname.  We scale the CDFs to emphasize tail latency.  The figure illustrates how both Clipper and INFaaS allow latency higher than their SLOs.  However, of note, with a 500\,ms SLO, INFaaS  successfully finds a configuration that can serve this SLO, and meets its SLO for 99\% of its requests.  By comparison, \sysname's tail latency remains very close to the SLO in all cases.  For the 500\,ms SLO, \sysname's latency remains at $\approx$300\,ms
because it schedules each model's entire batch of 16 requests at a time, round-robin across models.  With 15 models and a 20\,ms batch-16 execution duration, \sysname does not exceed the optimal 300\,ms latency.

\subsection{Can \sysname{} Serve Thousands?}
\label{sec:scaleupexperiment}

\begin{figure}%
\tikzsetnextfilename{bursty_experiment}%
\input{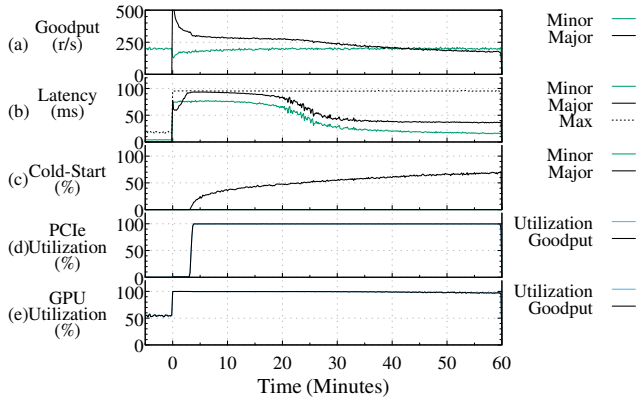}%
\vspace{-5mm}%
\caption{\sysname can serve thousands of models from a single worker.  From $t=0$, the Major workload adds an additional model per second, to a total of 3,600 models at $t=60$ (\cf~\autoref{sec:scaleupexperiment}.)}%
\label{fig:bursty_experiment}%
\end{figure}

\newcommand{\tequals}[1]{$t$\raisebox{.2ex}{$\mathsmaller{=}$}$#1$}

The previous experiment represented an idealized scenario, with only a small number of models, each with a steady sustained workload.  %
We now examine the serving limits of a single worker.  We deploy 3,601 copies of ResNet50 to a worker, and set a 100\,ms SLO.  We submit two workloads: a Major workload and a Minor workload.  The Major workload comprises 3,600 model instances; we vary the number of instances that are active at any point in time, and evenly distribute a workload of 1,000\,r/s across all active models.  The Minor workload is a single model instance that maintains a fixed 200\,r/s request rate throughout the experiment.

Figure~\autoref{fig:bursty_experiment} (a) plots the goodput achieved by the major and minor workloads.  From \tequals{-5} to \tequals{0} (we denote $t$ in minutes) only the Minor workload is present, achieving its full 200\,r/s.  At \tequals{0}, we activate one model instance of the Major workload; the addition of 1000\,r/s fully saturates the GPU (e). After that, we activate an additional model of the Major workload every 1 second.  As more model instances become active, the Major workload's goodput drops since each additional model forgoes batching opportunities.  At \tequals{60} all 3,600 models are active, each submitting approximately 0.28\,r/s.

By \tequals{3.5}, 201 models have been activated, reaching the capacity of GPU device memory.  To continue serving requests, \sysname begins swapping models on and off GPU; \autoref{fig:bursty_experiment} (d) shows PCIe utilization rapidly rises to 100\%.  As more models activate, an increasing number of requests in the Major workload find that their model is not loaded; \autoref{fig:bursty_experiment} (c) plots the rise in \emph{cold-starts}, reaching 70\% by the end of the experiment.  The minor workload, with its sustained request rate of 200\,r/s, does not experience any cold starts because its demand dwarfs every other model after the first 5 seconds.  As the number of cold-starts increases, the demand on GPU execution decreases, enabling the Minor workload's goodput to gradually grow back to 200\,r/s.  At approximately \tequals{20}, the bottleneck for the Major workload shifts to PCIe utilization, enabling the Minor workload's latency to drop back to an average of 20\,ms (b).

This experiment illustrates how bottlenecks in \sysname{} can shift as workload demand changes.  \sysname can deal with shifting bottlenecks even while serving a large number of models.  As illustrated in~\autoref{fig:bursty_experiment} (b), the maximum request latency across the experiment did not exceed the 100\,ms SLO.

\subsection{How Low Can \sysname Go?}
\label{sec:eval_slo_exp_1}
\sysname's predictability and centralized decision-making enables it to satisfy low-latency SLOs. %
In this experiment, we use six \sysname workers and evaluate the lower limit on SLOs that \sysname can achieve by measuring the proportion of successful requests while varying the SLO.
We repeat the experiment for six different workloads, varying the number of ResNet50 instances ($N=12$ or 48) and cumulative request rate ($R=600$\,r/s, 1200\,r/s, or 2400\,r/s).
For each experiment run, we begin with an SLO of 2.9\,ms (1$\times$ the execution latency of batch-1 ResNet50 inference).  Every 30 seconds, we extend the SLO by 50\%; by the end of the experiment the SLO reaches 74ms.
We run a separate open-loop client for each model with a Poisson inter-arrival time distribution, and as before, all models are independent (requests cannot be batched across models).

\begin{figure}%
\tikzsetnextfilename{slo_experiment_fig1}%
\input{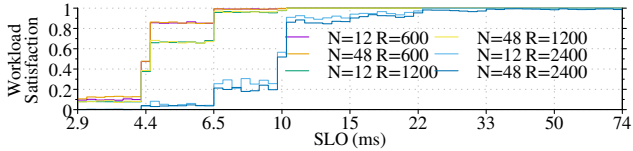}%
\vspace{-5mm}%
\caption{Workload satisfaction rates as we vary $N$, the number of clients, and $R$, the request rate.}%
\label{fig:clockwork-slo-exp-1}%
\end{figure}

\fakepara{Workload satisfaction.} \autoref{fig:clockwork-slo-exp-1} plots the \emph{workload satisfaction} for each experiment run. Workload satisfaction is the ratio of goodput to offered load.  A workload satisfaction of 1 means all requests received a successful response within their SLO.
For a load of $R=$600\,r/s and 1200\,r/s, irrespective of the number of models, \sysname{} successfully satisfied tight SLOs 10 and 22\,ms.
Even at $R=$2400~r/s, \sysname comfortably managed an SLO of 74ms.

\begin{figure}%
\tikzsetnextfilename{slo_experiment_fig2}%
\input{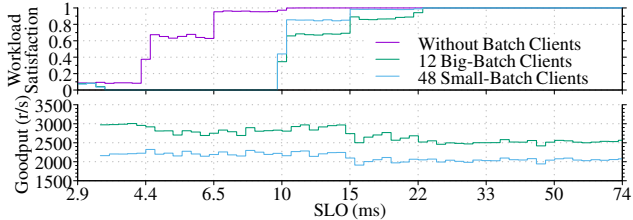}%
\vspace{-5mm}%
\caption{Workload satisfaction rates for latency-sensitive clients (top) and workload goodput for batch clients (bottom).}%
\label{fig:clockwork-slo-exp-2}%
\end{figure}

\subsection{Can \sysname{} Isolate Performance?}
\sysname{} can satisfy tight SLOs for latency-sensitive clients in isolation; we next consider when the system is shared with other users serving batch requests without latency SLOs.  As before, we use six \sysname workers, and all clients use instances of ResNet50. 
We provision six \emph{latency-sensitive} clients, each submitting a 200\,r/s open-loop workload.  We also provision several \emph{batch clients}, which submit sustained closed-loop workloads and do not have latency SLOs.  \emph{Big-batch} clients have a concurrency of 16, while \emph{small-batch} clients have a concurrency of 4.  Varying the concurrency affects the maximum batch size \sysname can achieve for batch client requests.  We considered three scenarios: \textbf{(a)}~baseline without batch clients; \textbf{(b)} 12 big-batch clients; and \textbf{(c)} 48 small-batch clients.

\autoref{fig:clockwork-slo-exp-2} illustrates the workload satisfaction rates for latency-sensitive clients and the total  goodput achieved for the batch clients.
\sysname{} successfully prioritizes latency-sensitive requests over batch requests. Through SLO-aware scheduling, it ensures that the workload satisfaction rates are unaffected by the presence of other pending, less time-critical requests.
At the same time, \sysname{} does not throttle batch requests entirely, but schedules them during idle times or expected idle times.
However, when the SLOs are too tight ($<$15ms), many latency-sensitive requests are rejected in advance, allowing pending batch requests to pass through.

\subsection{Are Realistic Workloads Predictable?}
\label{sec:azure_prediction_errors}

\begin{table}
\centering%
\newcommand{\thbc}[1]{\multicolumn{1}{c|}{\textbf{#1}}}%
\footnotesize%
\setlength\extrarowheight{0pt}%
\setlength\dashlinedash{\arrayrulewidth}%
\setlength\dashlinegap{1.5pt}%
\setlength\tabcolsep{5pt}%
\setlength\arrayrulewidth{0.3pt}%
\begin{tabular}{@{}lrp{1.8in}@{}}
\textbf{Model Family} & \textbf{Count} & \textbf{Model Variants} \\
\hline
DenseNet~\cite{huang2017densely} & 4 & 121, 161, 198, 201 \\
DLA~\cite{yu2018deep} & 1 & 34 \\
GoogLeNet~\cite{szegedy2015going} & 1 & \\
Inception~\cite{szegedy2016rethinking} & 1 & v3 \\
Xception~\cite{szegedy2016rethinking} & 1 & \\
MobilePose~\cite{howard2019searching} & 4 & SPRN18, MNv3, RN18, RN50 \\
ResNeSt~\cite{zhang2020resnest} & 4 & 14, 26, 40, 101 \\
ResNet~\cite{he2016deep} & 22 & 18, 18b, 34, 34b, 50, 50b, 50c, 50d, 50s, 50-1.8x, 101, 101b, 101c, 101d, 101s, 101-1.9x, 101-2.2x, 152, 152b,152c, 152d, 152s \\
ResNet-v2~\cite{he2016identity} & 5 & 18, 34, 50, 101, 152 \\
ResNeXt~\cite{xie2017aggregated} & 3 & 50-32, 101-32, 101-64 \\
SENet~\cite{hu2018squeeze} & 2 & 50-32, 101-32 \\
TSN~\cite{wang2016temporal} & 7 & iv1, iv3, r18, r34, r50, r101, r152 \\
Wide ResNet~\cite{zagoruyko2016wide} & 3 & 16-10, 28-10, 40-8 \\
Winograd~\cite{lavin2016fast} & 3 & RN18, RN50, RN101 \\
\end{tabular}%
\caption{List of models used in experiments.}
\label{table:allmodels}%
\end{table}

We now ask whether executions remain predictable  under realistic workloads that comprise many concurrent users and models. We also investigate whether \sysname{}  effectively exploits this predictability.

To answer these questions, we deploy \sysname on 12 workers and replay a workload trace of Microsoft Azure Functions (MAF)~\cite{shahrad2020serverless}.  The trace records approximately 46,000 function workloads, counting the number of invocations of each function, every minute, for two weeks.  It interleaves a wide range of workloads, including heavy sustained workloads, low utilization cold workloads, bursty workloads that fluctuate over time, and workloads with periodic spikes~\cite{shahrad2020serverless}.  We believe this to be a representative workload for evaluation since serverless platforms enable a wide range of applications and supporting ML inference on serverless is an active area of research~\cite{ishakian2018serving, bhattacharjee2019barista}.

In this experiment, we replay six hours of the MAF trace in real-time.  We use 61 different models (\autoref{table:allmodels}) 
taken from the ONNX Model Zoo~\cite{onnxmodelzoo} and the GluonCV Model Zoo~\cite{guo2020gluoncv}.  We duplicate each model 66 times, resulting in a total of 4,026 instances and reaching the main-memory capacity of our worker machines.  We replay ten or eleven function workloads for each model instance.  We configure \sysname with a 100\,ms SLO.

\begin{figure}%
\tikzsetnextfilename{azure_experiment_timeseries}%
\input{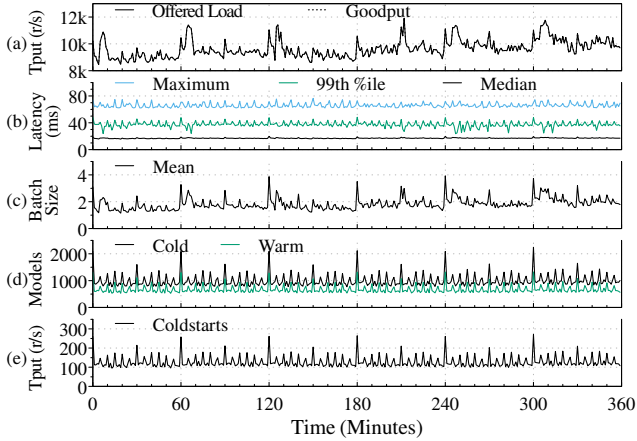}%
\vspace{-5mm}%
\caption{Microsoft Azure Functions (MAF) over \sysname{}; see~\autoref{sec:azure_prediction_errors} for a description.}%
\label{fig:azure_experiment}%
\end{figure}

\fakepara{\sysname with realistic workloads.}
The time series in \autoref{fig:azure_experiment} (a) shows the offered load and goodput achieved across all models.  For the 6 hour experiment, both the offered load and goodput averaged 9,638\,r/s -- out of a total of 208 million requests, only 58 failed due to action timing mispredictions, and no requests timed out.  All GPUs were fully utilized throughout the experiment, yet no request exceeded the 100ms SLO.

\autoref{fig:azure_experiment} (b) plots the median, \nth{99} percentile, and maximum request latency over the course of the experiment.  Latency spikes occur every 5, 15, and 60 minutes, due to the presence of numerous periodic workloads within the trace~\cite{shahrad2020serverless}.
Workload spikes do not cause SLO violations because of latency headroom; \autoref{fig:azure_experiment} (c) shows the average batch size for the experiment, and with each workload spike, \sysname can schedule larger batches, with higher latency.  To evaluate the cold-start behavior of this workload, we categorize a request as a cold-start if its model is not already loaded into GPU's memory before arrival.  For each 1-minute interval, \autoref{fig:azure_experiment} (d) counts the number of unique models that have at least one cold-start, and at least one warm-start.  On average, 987 unique models perform cold-starts each minute; or approximately 25\% of all models.  However, while many models perform cold-starts, they only represent a small fraction of all requests.  \autoref{fig:azure_experiment} (e) plots the throughput of cold-start requests, averaging 126\,r/s, or 1.3\% of all requests.
These results show that \sysname can sustain significant load for varied, realistic workloads comprising thousands of models.

\begin{figure}%
\tikzsetnextfilename{azure_experiment_cdf}%
\input{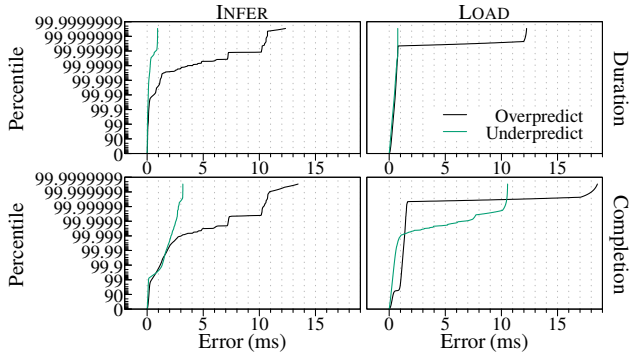}%
\vspace{-5mm}%
\caption{\sysname{} prediction and completion errors for MAF trace.}%
\label{fig:azure_experimenttail}%
\end{figure}

\fakepara{Predictable executions.}
\sysname's scheduler relies on accurate predictions of action latency, so to assess \sysname{}'s underlying assumptions of predictability, we next evaluate the accuracy of \sysname's predictions.  We measure the latency of \infer and \loadweights actions on \sysname's workers and compare it to the time estimated by \sysname's controller to derive a \emph{prediction error}.  Prediction errors comprise two types: \emph{overprediction}, when the real execution latency is faster than predicted; and \emph{underprediction}, when the real execution latency is slower than predicted.  
Consistent overpredictions can lead to idle resources, while consistent underpredictions can cause SLO violations.  
\autoref{fig:azure_experimenttail} (top) plots the prediction errors for \infer and \loadweights actions.  For \infer actions, the \nth{99} percentile of overpredictions and underpredictions is 144\,$\mu$s and 55\,$\mu$s, respectively.  Thereafter, the tail latency grows to exceed  10\,ms in a few extremely rare cases.  \sysname consistently overpredicts more than it underpredicts, as it uses a rolling \nth{99} percentile measurement to make its predictions.  For \loadweights actions, the \nth{99} percentile of overpredictions and underpredictions is 431\,$\mu$s and 348\,$\mu$s, respectively.  

\autoref{fig:azure_experimenttail} (bottom) plots the \emph{completion time error}.  \sysname must accurately predict when a given action will complete, taking into account any previously submitted actions (\ie queuing time).  Individual prediction errors can compound, leading to increased completion time error.  For \infer actions, the error compounds $4\times$, with a \nth{99} percentile completion error of $\approx$1\,ms.  In extreme cases, \sysname's completion error also grows to more than 10\,ms.  However, the completion error does not substantially exceed the action duration error, implying that for \sysname,  erroneous predictions of outliers are statistically independent.

\begin{figure}%
\tikzsetnextfilename{scalability_experiment}%
\input{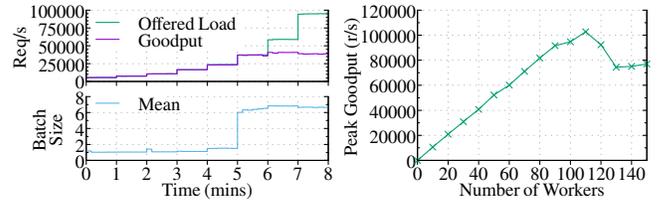}%
\caption{(Left) With 40 emulated workers, goodput is approximately equal to offered load; peak goodput is achieved at appx. 40,000\,r/s, when all workers are fully utilized.  (Right) Peak goodput achieved with different numbers of emulated workers.}
\label{fig:clockwork-scalability-exp}
\end{figure}

\subsection{Can Clockwork Scale?}
\label{sec:scalabilityexp}
Centralized scheduling presents a potential scalability bottleneck, though prior work has demonstrated that centralized schedulers can reach impressive scale~\cite{gog2016firmament, perry2014fastpass}.  Our final experiment examines the scalability of Clockwork's controller.

To venture beyond the capacity of our testbed, we leverage a specially-developed \emph{emulated worker} that implements Clockwork's action interface.  The emulated worker behaves identically to a bona fide Clockwork worker, except the \loadweights and \infer actions perform no meaningful work; instead, they wait for a period of time according to the pre-profiled model measurements before returning a response.  
The emulated worker is indistinguishable from a real worker from the vantage point of Clockwork's controller.  To bypass the limited network capacity of our testbed, we modified our clients to send zero-length inputs (network is not a fundamental limitation; see~\autoref{sec:discussion} for discussion).

We measure the peak goodput achieved as we vary $N$, the number of emulated workers.  We run multiple experiments, each with a different value of $N$, from 10 to 150 in increments of 10.  We use the same models as described in~\autoref{sec:azure_prediction_errors}, and a similar workload.  Instead of replaying the trace at a fixed rate, we scale the trace and gradually offer more load in 60-second intervals.  \autoref{fig:clockwork-scalability-exp} (Left) illustrates one experiment run with $N$=40.  Goodput follows the offered load almost perfectly up to about 40,000\,r/s, at which point all workers are fully utilized and the goodput saturates.

\autoref{fig:clockwork-scalability-exp} (Right) reports the peak goodput achieved with different numbers of workers.  We report the median values across three experiment repetitions.  The figure shows a linear increase in the peak goodput as the number of workers increases.  Below $N$=110, goodput is limited by workers reaching full utilization.  At $N$=110, we reach a maximum goodput of 103,387\,r/s.  At this point worker utilization stops being the limiting factor; instead, the bottleneck shifts to Clockwork's controller.  Beyond $N$=110 peak goodput declines.

\subsection{Summary}

In comparison with prior model serving systems, \sysname achieves superior goodput, serves considerably more models concurrently, and violates substantially fewer SLOs.
Owing to a lack of performance variability, \sysname{} can achieve much tighter latency SLOs without sacrificing tail latency.
\sysname{}'s underlying assumptions about predictable executions bear out in reality: by consolidating choice, a predictable system that substantially curtails tail latency can be built. 

\sysname extends to a diverse range of workload conditions not supported by prior systems, including supporting thousands of models concurrently per GPU.  Slow cold starts can run alongside high-throughput workloads without interference.
Under all workload conditions, including cold starts and even under overload, \sysname meets most SLOs %
without degrading service, and maintains close to maximal possible goodput. 
Finally, \sysname{} isolates users of different models, enabling low-latency workloads to share the same system with background batch workloads.

\section{Discussion}
\label{sec:discussion}

\fakepara{Why consolidate choice?}
Philosophically, the encapsulation, abstraction, and loose coupling of components are essential design practices while the building blocks and use cases of large systems are still in flux. 
Over time, the true use cases for the system settle and the entire system may in turn be replaced by a simpler, refined system that avoids the over-engineering and generality of its constituent parts---components that transpired to either be unnecessary in practice or to impede the common use case of the system. 
The squashing of layers through such specialization, effectively transforming systems into abstract units, can counteract the infamous bloat of modern software stacks. 
We designed \sysname{} to be such an abstract unit for model serving systems.

\fakepara{Machine learning.}
\sysname{} focuses on DNN inference, and excludes data preprocessing and postprocessing steps that are  user-defined and CPU-bound. Safely and predictably executing these in \sysname{} is a current research topic.

Individual DNN inferences are the atomic unit of work for Clockwork.
Increasingly, modern ML applications are composed of pipelines or cascades of
DNNs~\cite{hsieh2018focus, kang2017noscope,
shen2019nexus}.  For these applications, performance predictability is strongly
desired.
We believe there are opportunities to leverage Clockwork's properties
and perform more sophisticated pipeline scheduling that provides end-to-end
guarantees.
Similarly, performance predictability can influence system designs in other areas, such as large language embedding models~\cite{brown2020language} that may require dedicated or distributed accelerators.
Expanding \sysname{} into other ML paradigms, such as deep reinforcement learning and DNN training, raises 
philosophical questions about the nature and limits of predictability.

\fakepara{Inference accelerators.}
The \sysname{} approach generalizes readily beyond GPUs to other inference-specific hardware accelerators~\cite{mattson2020mlperf}, whose performance is arguably even more predictable.
TPUs~\cite{jouppi2017datacenter}, for instance, are explicitly built around the idea of delegating control to software,
while also 
eschewing
general purpose processing engines with flexible control logic and generic memory hierarchies in favor of high-level operations and explicit memory hierarchies.

On the other extreme, inferences can also be executed in software on the CPU.
While many models are heavily parallel in nature and execute orders of magnitude slower on CPUs, there are other models where execution on CPU is acceptable.
One such example are recurrent neural networks (RNNs) which are fundamentally more sequential and often cannot effectively leverage the available parallelism on GPUs or other accelerators.

\fakepara{Limitations of predictability.}
Consolidating choice is only possible when you have control of, or guarantees about, the system's major bottleneck resources.  For example, Clockwork assumes workers have exclusive control over their machine, and dedicated GPUs.  Clockwork does not assume exclusive control over the network, but does assume that the network has mostly-predictable latency between the controller and workers.  In a shared setting, preserving predictability becomes more challenging -- though not impossible -- and this is an active area of research due to a general need to co-locate latency critical datacenter services~\cite{lo2015heracles, kaffes2020leveraging}.

\fakepara{Network.}
\label{sec:discussion-net}
\sysname{} does not explicitly consider the network in its scheduling decisions; 
the occasional network latency spikes of dozens of ms during our experiments had negligible
impact on our results. %
Our prototype routes all inputs and outputs through the central controller which will become a bottleneck at scale.  We were able to reach the limits of our testbed network with 12 workers and a sustained, single-model workload; to test beyond this we disabled inputs as described in~\autoref{sec:scalabilityexp}.  This limitation is not fundamental; Clockwork's controller only requires request \emph{metadata} to schedule requests, and we are working to remove this limitation with a tier of load balancers.

\fakepara{Security.}
Security is important for all multi-user systems, since there are no container or hypervisor boundaries separating the workloads of different users.
\sysname{} does not explicitly address security; however, \sysname{} does not execute arbitrary user code.  Users must submit models in an abstract format that we then compile to binary code under the covers.  \sysname{}'s threat model resembles shared storage or database systems, where system correctness is the chief concern; we have not verified any safety properties of \sysname{}.

\fakepara{Fault tolerance.}
While \sysname{} is a distributed system, we do not address the challenges of tolerating failures when serving models at large scale.%
This will require implementing a fault-tolerant centralized scheduler; however, we note that \sysname{}'s predictable worker design will make pernicious phenomena like grey failure~\cite{gunawi2018fail,huang2017gray} far easier to detect.

\fakepara{Other benefits of predictability.}
Concentrating choice makes it easier to implement other guarantees, such as SLOs related to burstiness or per-request cost.  The Azure trace in our evaluation, for instance, contained regular, periodic spikes; exploiting advanced knowledge is an appealing future avenue for \sysname{}.  A further benefit of predictable system components is \emph{performance clarity}~\cite{ousterhout2017monotasks}: performance bottlenecks and upcoming tasks in \sysname{} are easy to reason about.  \sysname{}'s controller also provides a central point for \emph{explanation}, since the controller has complete visibility of the expected and actual request behavior.

\section{Related Work}

\fakepara{Model serving.}
We directly compared \sysname{} to
Clipper~\cite{crankshaw2017clipper} and INFaaS~\cite{romero2019infaas} in~\autoref{sec:evaluation-compare}; here we
provide additional comments. Both Clipper and INFaaS are designed as
wrappers around existing model execution frameworks: Clipper, in order to
provide a unifying abstraction; INFaaS, in order to exploit heterogeneous
execution strategies.  Being agnostic to the underlying execution engine
sacrifices predictability and control over model execution.  Both systems treat
latency SLOs as long-term, reactive targets; by contrast, Clockwork is
explicitly designed to consolidate choice, and exploit predictability by making
proactive decisions.  Clipper and INFaaS propose several orthogonal concepts
that are compatible with Clockwork.  Clipper's model selection layer could be
superimposed on Clockwork.   INFaaS's model variant concept could be integrated
into Clockwork; we found similar predictability properties held for DNNs
executing on dedicated CPU cores.  

Several other projects investigate model serving in virtualized cloud
environments and on serverless platforms,
where predictability is in the hands of the cloud provider
\cite{zhang2019mark,bhattacharjee2019barista,kannan2019grandslam}.
Like INFaaS, these model throughput, latency, and accuracy together for
optimal model selection, but, unlike \sysname{}, they do not use the backend
predictability and latency SLOs for making proactive scheduling decisions.
In industry, TFS$^2$ \cite{olston2017tensorflowserving} is a proprietary model
hosting service at Google, about which public information is not available.  
Amazon SageMaker~\cite{sagemaker} and Google AI
Platform~\cite{googleaiplatform} are public cloud DNN serving systems with a
similar interface to Clockwork: upload your model, then make inference
requests.  Both use containers under the covers as an isolation mechanism, and
users suffer the associated cold-start latency.  Beyond these details, further
design information is not publicly known.

\fakepara{Real-time systems.}
Performance predictability, especially temporal safety,
is also an important concern for safety-critical real-time systems.
However in general, real-time systems are designed for \textit{periodic} or
\textit{sporadic} workloads~\cite{baker2007schedulability} with known minimum
inter-arrival times and worst-case execution times, or for scenarios
where the set of all inference requests is known in advance~\cite{sun2020real}.
\textit{Soft-real-time} systems~\cite{buttazzo2005soft}  consider
weaker notions of timeliness similar to the latency SLOs considered in this paper, but mainly target periodic or sporadic workloads.
Clockwork, in contrast, makes no \apriori{} assumptions about its workloads.
Prior real-time systems work has also proposed mechanisms
to tame the unpredictability inside GPUs
\cite{elliott2011real,elliott2013optimal,amert2017gpu,otterness2017inferring,bakita2018scaling}.
Elliott and Anderson~\cite{elliott2012robust}, for example, proposed interrupt
handling mechanisms to circumvent the proprietary GPU drivers that ignore
scheduling priorities, while Yang~\emph{et~al.}~\cite{yang2018avoiding}  suggested avoiding synchronization anomalies through more careful use of
CUDA synchronization primitives.
These mechanisms are designed to facilitate an \apriori
\textit{schedulability analysis}---
mathematically bounding the blocking delays due to contention. Such bounds are orthogonal to Clockwork, which does not require strict worst-case guarantees.

\section{Conclusion}

As DNN inferences become increasingly central to interactive applications, the requirements for fast response tighten, the volume of requests expands, and the number of models grows.
Our model serving system, \sysname{}, meets these challenges.
\sysname{} efficiently fulfills aggressive tail-latency SLOs
while supporting thousands of DNN models with different workload characteristics concurrently on each GPU, and scaling out to additional worker machines for increased capacity.
The system also successfully isolates models from performance interference caused by other models served on the same system.
Our results derive from our design methodology of recursively ensuring all internal architecture components have predictable performance by concentrating all choices in the centralized controller. 
Notably, our approach required us to either circumvent canonical best-effort mechanisms or orchestrate them to become predictable,
and illustrates how consolidating choice can be applied in practice to achieve predictable performance.

\subsection*{Acknowledgements}
We thank our shepherd Junfeng Yang and the anonymous reviewers for their insightful 
feedback that helped improve our work. 
Our work was partially supported by NSF CAREER Grant \#1553579.

\bibliographystyle{plain}
\bibliography{ref}

\appendix
\clearpage

\appendix

\section{Artifact Appendix}

\subsection{Abstract}

The artifact consists of Clockwork's prototype source code, instructions for building from source, and directions for preparing the environment. The instructions for launching a Docker instance that has all dependencies pre-installed is provided as well. The artifact also contains scripts, descriptions, and instructions to run the experiments automatically or manually for reproducing the graphs and results presented in the paper.

\subsection{Artifact check-list}

{\small
\begin{itemize}[leftmargin=*,label=$\cdot$]
  \item {\bf Program: } dnn-model-serving, multi-tenant
  \item {\bf Compilation: } cmake, g++
  \item {\bf Binary: } worker, controller, client
  \item {\bf Model: } distributed, multi-tenant
  \item {\bf Data set: } azure-functions-trace-2019, poission-distribution
  \item {\bf Run-time environment: } Linux, CUDA, network
  \item {\bf Hardware: } NVIDIA, Tesla-V100
  \item {\bf Execution: } automated, manual
  \item {\bf Metrics: } throughput, latency, SLO-violation, tail-latency
  \item {\bf Output: } telemetry-measurements, table, graph
  \item {\bf Experiments: } throughput-latency, scalability, predictability, SLO, tail-latency
  \item {\bf Required disk space: }\\
  Clockwork: 210MB\\
  Total including compiled models and dataset: 12GB
  \item {\bf Expected experiment run time:} About 17~hours in total
  \item {\bf Public link:} \\
  \url{https://gitlab.mpi-sws.org/cld/ml/clockwork}
  \item {\bf Code licenses: }\\
  Clockwork: Apache License 2.0\\
  TVM: Apache License 2.0\\
  CUDA Common Library: Apache License 2.0\\
  Catch2: Boost Software License 1.0
  
  \item {\bf Data licenses: }\\
 Azure Functions Trace 2019: CC-BY Attribution
\end{itemize}
}
\subsection{Description}

\subsubsection{How to access}
The artifact is publicly available at \\
\url{https://gitlab.mpi-sws.org/cld/ml/clockwork}

\subsubsection{Hardware dependencies}
To reproduce the exact experiment results, worker machines must have 768GB RAM or higher, 16 CPU cores or more, at least one 32GB Tesla v100 GPU and 10Gbps network. The large-scale experiment with Azure Functions (\autoref{fig:azure_experiment}) requires 12 worker machines. Most other experiments require fewer worker machines; details on the number of machines for each experiment and environment customization guide is provided in each experiment's documentation.

\subsubsection{Software dependencies}
\begin{itemize}[leftmargin=*,label=$\cdot$]
\item \textbf{Clockwork:} \\
Ubuntu 18.04 or later, CUDA v9.0+, libtbb-dev, libasio-dev, libconfig++-dev, libboost-all-dev, g++-8, 
make, cmake, automake, autoconf, libtool, curl, unzip, clang, llvm, and protobuf.

A Dockerfile is provided to facilitate the build process.

\item \textbf{Data analysis and plotting scripts:} \\
Python 3.x and the numpy, pandas, matplotlib, and seaborn libraries.

\end{itemize}
\subsubsection{Data sets}
\begin{itemize}[leftmargin=*,label=$\cdot$]
    \item Publicly released Azure Functions 2019 trace~\cite{shahrad2020serverless}\\ \url{https://gitlab.mpi-sws.org/cld/trace-datasets/azure-functions}
\end{itemize}
\subsubsection{Models}
The DNN models pre-compiled for NVIDIA Volta V100 GPUs are accessible at\\ \url{https://gitlab.mpi-sws.org/cld/ml/clockwork-modelzoo-volta}
\subsection{Installation}
\begin{itemize}[leftmargin=*,label=$\cdot$]
    
\item Installation pre-requisites:\\
\url{https://gitlab.mpi-sws.org/cld/ml/clockwork/-/blob/master/docs/prerequisites.md}

\item Building Clockwork:\\
\url{https://gitlab.mpi-sws.org/cld/ml/clockwork/-/blob/master/docs/building.md}

\item Setting-up the environment:\\
\url{https://gitlab.mpi-sws.org/cld/ml/clockwork/-/blob/master/docs/environment.md}

\item Clockwork configuration:\\
\url{https://gitlab.mpi-sws.org/cld/ml/clockwork/-/blob/master/docs/configuration.md}

\end{itemize}

\subsection{Experiment workflow}

Experiments can be run using the scripts provided in the repository.  We have also provided instructions to run the experiments manually.
To get started with Clockwork, we recommend getting the system running manually, in order to understand the pieces involved, and to ensure the system has been configured appropriately for your machines.  Afterwards, you might choose to run the experiments using the provided scripts or manually. The experiments repository is available at\\
\url{https://gitlab.mpi-sws.org/cld/ml/clockwork-results}

\subsection{Evaluation and expected results}

\begin{table*}[t]
\centering%
\newcommand{\thbc}[1]{\multicolumn{1}{c|}{\textbf{#1}}}%
\newcommand{\thbl}[1]{\multicolumn{1}{l|}{\textbf{#1}}}%
\newcommand{\trbc}[1]{\multicolumn{1}{c|}{#1}}%
\footnotesize%
\setlength\extrarowheight{1.5pt}%
\setlength\dashlinedash{\arrayrulewidth}%
\setlength\dashlinegap{1.5pt}%
\setlength\tabcolsep{5pt}%
\setlength\arrayrulewidth{0.3pt}%
\begin{tabular}{m{0.33\textwidth}|m{0.1\textwidth}|r|m{0.4\textwidth}}
  \multirow{2}{*}{\textbf{Experiment}} &
  \multirow{2}{*}{\textbf{Related figure}} &
  \multirow{2}{*}{\textbf{Execution}} &
  \multirow{2}{*}{\textbf{Documentation and scripts}} \\
  & & \thbl{time (hr)} & \\
  
\hline

\multirow{1}{*}{How Does Clockwork Compare?}
& \autoref{fig:simple_experiment} & \trbc{3} & \url{https://gitlab.mpi-sws.org/cld/ml/clockwork-results/-/tree/master/sec61_fig5} \\
\cdashline{1-4}
\multirow{1}{*}{Can Clockwork Serve Thousands?}
& \autoref{fig:bursty_experiment} & \trbc{1.5}  & \url{https://gitlab.mpi-sws.org/cld/ml/clockwork-results/-/tree/master/sec62_fig6} \\
\cdashline{1-4}
\multirow{1}{*}{How Low Can Clockwork Go?}
& \autoref{fig:clockwork-slo-exp-1} & \trbc{1}  & \url{https://gitlab.mpi-sws.org/cld/ml/clockwork-results/-/tree/master/sec63_fig7} \\
\cdashline{1-4}
\multirow{1}{*}{Can Clockwork Isolate Performance?}
& \autoref{fig:clockwork-slo-exp-2} & \trbc{1} & \url{https://gitlab.mpi-sws.org/cld/ml/clockwork-results/-/tree/master/sec64_fig8} \\
\cdashline{1-4}
\multirow{1}{*}{Are Realistic Workloads Predictable? }
& \autoref{fig:azure_experimenttail}  & \trbc{8}  & \url{https://gitlab.mpi-sws.org/cld/ml/clockwork-results/-/tree/master/sec65_fig9_fig10} \\
\cdashline{1-4}
\multirow{1}{*}{Can Clockwork Scale?}
& \autoref{fig:clockwork-scalability-exp} & \trbc{2} &  \url{https://gitlab.mpi-sws.org/cld/ml/clockwork-results/-/tree/master/sec66_fig11} \\

\end{tabular}%
\caption{The experiments reproducing the presented results in this paper, their related figures, execution time, and links to the extensive documentation and scripts for each experiment. }      
\label{table:aeexperiments}
\end{table*}

The experiments repository is structured based on \autoref{sec:evaluation}. We have provided the experiment titles, their related figures on the paper, execution time of each experiment, and the links to directories containing the respective descriptions, scripts and instructions in \autoref{table:aeexperiments}.

\subsection{Experiment customization}

The directions for running each experiment manually is provided in each experiment's documentation. Instructions for customizing the experiment environment is provided at\\
\url{https://gitlab.mpi-sws.org/cld/ml/clockwork/-/blob/master/docs/customizing.md}

\subsection{AE Methodology}

Submission, reviewing and badging methodology:\\
\url{https://www.usenix.org/conference/osdi20/call-for-artifacts}

\clearpage

\end{document}